\documentclass[amsfonts,amsmath,amssymb,prb,twocolumn,showpacs]{revtex4}
\usepackage{graphicx}
\usepackage{color}
\usepackage{bm}
\usepackage{bbm}
\usepackage{dsfont}
\usepackage{mathbbol}
\newcommand{\vectornorm}[1]{\left|\left|#1\right|\right|}
\newcommand{\einheit}[1]{\hspace{1mm}\mbox{#1}}
\newsavebox{\mysquare}
\savebox{\mysquare}{\textcolor{black}{\rule{2.5mm}{2.5mm}}}
\usepackage{float}

\begin{document}

\title{Bell--state preparation for electron spins in a semiconductor double quantum dot}
\author{Robert Roloff}
\email{robert.roloff@edu.uni-graz.at}
\author{Walter P\"{o}tz}
\email{walter.poetz@uni-graz.at}
\affiliation{Fachbereich Theoretische Physik, Institut f\"{u}r Physik, Karl-Franzens-Universit\"{a}t Graz, Universit\"{a}tsplatz 5, 8010 Graz, Austria}
\date{\today}
\begin{abstract}
A robust scheme for state preparation and state trapping for the spins of two electrons in a semiconductor double quantum dot is presented.  The system is modeled by two spins coupled to two independent bosonic reservoirs. Decoherence effects due to this environment are minimized by application of optimized  control fields which make the target state to the ground state of the isolated driven spin system. We show that stable spin entanglement with respect to pure dephasing is possible. Specifically, we demonstrate state trapping in a maximally entangled state (Bell state) in the presence of decoherence.
\end{abstract}

\maketitle

\section{Introduction}\label{sec:intro}
In recent years, there has been a great interest in quantum computing and quantum information processing. Since Shor demonstrated that his quantum algorithm is able to perform prime factorization more efficiently than any known classical algorithm, 
the attention paid to this type of computation has grown enormously.~\cite{Niel02,Shor97}\par
Within a network model, the fundamental building block  of a quantum computer is the \textit{quantum bit} or \textit{qubit}. A two--level system (TLS) which can provide coherent state superposition is needed to physically realize a quantum bit. There are several suggestions in the literature for achieving the goal of finding a suitable TLS. Amongst such proposed systems are quantum dots,~\cite{DiVi98} single photons,~\cite{Yor03} the nuclear spin,~\cite{Ger97} superconducting devices,~\cite{Orl99, Mak99} and trapped ions.~\cite{Cir95}\par
Solid state realizations of qubits and quantum gates are attractive because they may be quite easily scaled to larger arrays of qubits which is necessary for efficient quantum computing. A major advantage of the semiconductor solid state implementation is that the existing semiconductor technology is well developed. Because of these advantages, we are going to present our approach in terms of semiconductor spin quantum dots, although the model and strategies developed here can be mapped onto most qubit realizations which have been discussed in the literature.\par
Quantum computation is powerful because it takes advantage of quantum interference ({\it i.e.}, coherence) and entanglement.  However, when a quantum system is in contact with an environment, the resulting interaction destroys coherence and changes the populations. Several methods 
for suppression or avoidance of (unwanted effects of) decoherence, such as  quantum error correction,~\cite{Niel02,Pres98} encoding of the qubit into a decoherence free subspace,~\cite{Lid98,Lid99} and dynamical decoupling,~\cite{Viol98,Viol99} have been proposed. Another technique to cope with dissipation is to make the desired state to the ground state of the system by tuning of (possibly adiabatic) control fields.~\cite{Poetz1}
\par 
In this paper, we study the feasibility of Bell--state generation for two electron spins in a semiconductor double quantum dot.  The present work is structured as follows. In the first part, we outline the model which is used to describe the system of interest. The second part is devoted to the derivation of the equations of motion and the calculation of the system--bath correlation functions. In the next part, the strategy for fighting decoherence is outlined and numerical results are given. We will end with summary and conclusions of the present work.

\section{Theory}\label{sec:theory}

The spin $\frac{1}{2}$ of an electron provides a natural two--level system.  An electron in a semiconductor quantum dot as a realization of a qubit has been proposed by Loss and DiVincenzo.~\cite{DiVi98} The spin directions \textit{up} and \textit{down} with respect to an external magnetic field represent the two basis states of the qubit. Few-electron semiconductor quantum dots can be realized, for example, by means of surface gates on top of a GaAs/AlGaAs heterostructure which holds a two dimensional electron gas.~\cite{Elz} The number of conducting electrons in each dot can be controlled and monitored.~\cite{Elz}\par
It has been shown that a system consisting of two neighboring quantum dots, each populated by one excess  electron, can be described by a Heisenberg--like Hamiltonian $H_S^{(12)}=J(t)\vec S_1 \cdot \vec S_2$ if the dots are coupled via a tunable tunneling barrier.~\cite{Burk99,DiVi98} The straightforward way to implement such a two--qubit array is to use a heterostructure as described above and to increase the number of gate electrodes.~\cite{Elz} In a recent paper it has been argued that for quantum dots (which are spatially separated so that tunneling between the dots can be neglected), controlled entanglement via electrostatic interaction is possible.~\cite{Trif07}
\par
First, the Hamilton operators for the double quantum dot, the environment, and the interactions on which our investigation is based are defined. To include the influence of the environment in the dynamics of the double dot, a Markovian quantum master equation approach is used below.~\cite{Car02,Breu03,Scully} The system--bath correlation functions are calculated analytically.

\subsection{Description of the system}\label{sec:theory-description}

The model system consists of two conduction electrons, each sitting in its own quantum dot. Each is described by a Hamiltonian of the form
\begin{eqnarray}
H_S^{(i)}(t)&=&-g^{*}\frac{e}{2m_e^{*}} S_z B^{(i)}_z(t) \nonumber \\
&=&-\frac{g^{*}}{2}\mu_B B^{(i)}_z(t) \sigma_z^{(i)} \nonumber \\ 
&\equiv&-\hbar \tilde B^{(i)}_z(t) \sigma_z^{(i)},\quad i=1,2 \label{H_S} \\
\vec S^{(i)}&=& \frac{\hbar}{2} \vec \sigma^{(i)}.
\end{eqnarray}
$g^{*}$ denotes the gyromagnetic ratio which depends, as does $m_e^{*}$, on the type of the semiconductor quantum dot.
This Hamilton operator describes the interaction of the $i$th electron spin with an external magnetic field applied in the $z$ direction. $B^{(i)}_z(t)$ is a control field which is used to adjust the Zeeman splitting associated with the electron spin in quantum dot $i$. For the most general case of magnetic fields, there are nonzero $x$ and $y$ components, e.g., to perform single qubit operations on the spin. \par
We model the presence of an environment which is responsible for dephasing by coupling the electron spin degree of freedom to a bath of bosons. Such models have been widely used in the literature to describe various forms of interactions at a microscopic level by selection of the spectral properties of the bosonic bath. In the present system, the main cause for dephasing arises from charge fluctuations in the vicinity of the quantum dots, phonons, and interaction with nuclear spins.~\cite{Tay06,Gol04,Moz02,Yu02,Hu06}
The influence of the phonons on the spin degree of freedom is mediated through the spin--orbit interaction. At low temperature, acoustic electron--phonon interaction becomes the main scattering mechanism.~\cite{Mit99} Therefore, we can assume that only LA and TA phonons are involved in the scattering process. As all nuclear spins corresponding to $\mbox{III-V}$ semiconductor materials, such as GaAs, have nonzero nuclear spins, the hyperfine interaction of the latter with the electron spin is another source of decoherence.~\cite{Tay06} Moreover, for exchange--interaction--based quantum gates, charge fluctuations produced by tuning the gate voltages lead to dephasing.~\cite{Hu06}\par
We model the main effects from these interactions, {\it i.e.}, dephasing, using two uncorrelated baths of harmonic oscillators, 
\begin{equation}
{H_R}^{(i)} = \sum\limits_k{\hbar \omega_k^{(i)} b_k^{(i)\dag}b_k^{(i)}}, \; i=1,2, \label{H_R}
\end{equation}
where $b_k^{(i)\dag}$ and $b_k^{(i)}$ are the bosonic creation and annihilation operators for modes with frequency $\omega_k^{(i)}$.
The interaction of the baths with the qubits is of spin-boson type,~\cite{Leg87,Weiss99}
\begin{eqnarray}
& &H_{SR}^{(i)}=\hbar \sigma_z^{(i)}\Gamma^{(i)}, \label{H_SR}\\
& &\Gamma^{(i)}=\sum\limits_k{g_k^{(i)}\left( b_k^{(i)}+b_k^{(i)\dag}\right). \label{gamma_def}}
\end{eqnarray}
The $g_k^{(i)}$ are the effective coupling constants of the spin--boson interaction.
For the case of strongly correlated reservoirs (collective decoherence), the decoherence free subspace includes a set of maximally entangled states between the two spins (see Sec.~\ref{sec:theory-eom}). In Ref.~\onlinecite{Yu02}, it has  been shown that for the case of diagonal qubit--qubit coupling and collective decoherence a set of robust entangled states can be identified. So, here we examine the more challenging (and probably more realistic) case of two independent reservoirs. In addition, we are going to implement the full Heisenberg interaction between the two spins.\par The coupling to $\sigma_z^{(i)}$ in Eq.~\eqref{H_SR} leads to pure dephasing, {\it i.e.}, destruction of coherence. Terms proportional to $\sigma_x^{(i)}$ or $\sigma_y^{(i)}$ would cause population relaxation which, within this model, corresponds to spin flips. Spin flips do occur in semiconductors. However, these processes in general take 
place on a much larger time scale than dephasing processes, {\it i.e.}, the spin relaxation time $T_1$ is much longer than the dephasing time $T_2$, and so relaxation is neglected here. In this sense we study the quantum dots on a time scale $t< T_1$.\par

The qubit--qubit interaction which is needed to produce entanglement and conditional operations is of Heisenberg type,~\cite{DiVi98,Burk99}
\begin{eqnarray}
H_S^{(12)}(t) &=& J(t)\vec\sigma^{(1)}\cdot \vec\sigma^{(2)}, \label{H_S_12}\\
\vec\sigma^{(1)}\cdot \vec\sigma^{(2)} &\equiv & \sigma_x^{(1)}\otimes\sigma_x^{(2)}+\sigma_y^{(1)}\otimes\sigma_y^{(2)}+\sigma_z^{(1)}\otimes\sigma_z^{(2)}\label{def_otimes}.\nonumber
\end{eqnarray}
The Hamiltonian of the externally controlled two--qubit system is
\begin{eqnarray}
& &H_S(t)={H_S}^{(1)}(t)\otimes \openone+\openone \otimes {H_S}^{(2)}(t)+{H_S}^{(12)}(t) \label{H_S_ges} \\
& &=-\hbar \tilde B^{(1)}_z(t) \sigma_z^{(1)}\otimes \openone-\hbar \tilde B^{(2)}_z(t) \openone \otimes \sigma_z^{(2)}+J(t)\vec\sigma^{(1)}\cdot \vec\sigma^{(2)}. \nonumber 
\end{eqnarray}
The system-reservoir Hamiltonian to introduce the interaction between the environment and the system reads
\begin{eqnarray}
H_{SR}&=& H_{SR}^{(1)}+H_{SR}^{(2)} \nonumber\\
&=& \hbar\left( \sigma_z^{(1)} \otimes \openone \right) \Gamma^{(1)}+\hbar\left( \openone \otimes \sigma_z^{(2)}\right) \Gamma^{(2)},\label{H_SR_ges}
\end{eqnarray}
and the reservoir Hamiltonian consisting of the two independent baths is of the form
\begin{eqnarray}
H_{R}&=&H_{R}^{(1)}+H_{R}^{(2)} \nonumber\\
&=& \sum\limits_k{\hbar \omega_k^{(1)} b_k^{(1)\dag}b_k^{(1)}}+\sum\limits_k{\hbar \omega_k^{(2)} b_k^{(2)\dag}b_k^{(2)}}. \label{H_R_ges}
\end{eqnarray}
The full Hamiltonian of the interacting system is
\begin{equation}
H(t) = H_{S}(t)+H_{R}+H_{SR}.\label{H_ges}
\end{equation}
The representation of $H(t)$ in the system space is of block--diagonal form and we separate it into 
an ``outer" and an ``inner" contribution, respectively, $H_a(t)$ and $H_b(t)$ 
\begin{eqnarray}
H(t)&\equiv&H_a(t)+H_b(t)=\begin{bmatrix}
H_{11}(t) & 0 & 0 & 0 \\
0 & 0 & 0 & 0 \\
0 & 0 & 0 & 0 \\
0 & 0 & 0 & H_{44}(t)
\end{bmatrix}\nonumber \\
&+&\begin{bmatrix}
0 & 0 & 0 & 0 \\
0 & H_{22}(t) & H_{23}(t) & 0 \\
0 & H_{23}^*(t) & H_{33}(t) & 0 \\
0 & 0 & 0 & 0 
\end{bmatrix}.\label{H_S_1}
\end{eqnarray}
With $|0\rangle $ and $|1\rangle$, respectively, denoting spins down and up, we use the following convention for the four basis states $|i\rangle \otimes |j\rangle \equiv |ij\rangle,\;\mbox{for}\quad i,j=0,1$:  
\begin{eqnarray}
& &\left.|1\right)  \equiv  |11\rangle \doteq \left( 1,0,0,0\right)^T,\; \left.|2\right)\equiv |10\rangle \doteq \left( 0,1,0,0\right)^T, \nonumber \\
& &\left.|3\right)  \equiv  |01\rangle \doteq \left( 0,0,1,0\right)^T,\; \left.|4\right)\equiv |00\rangle \doteq \left( 0,0,0,1\right)^T, \nonumber \\
& &A = \sum_{i,j=1,2,3,4} A_{ij}\left.|i\right) \left( j|\right. \nonumber
\end{eqnarray}
for any operator $A$ defined on the Hilbert space of the double--qubit system.
One then finds
\begin{eqnarray}
& &H_{11}(t)=-E_1(t)-E_2(t)+J(t)+\hbar \Gamma^{(1)} +\hbar \Gamma^{(2)} +H_R, \nonumber\\
& &H_{22}(t)=-E_1(t)+E_2(t)-J(t)+\hbar \Gamma^{(1)} - \hbar \Gamma^{(2)} +H_R, \nonumber \\
& &H_{33}(t)=E_1(t)-E_2(t)-J(t)-\hbar \Gamma^{(1)} + \hbar \Gamma^{(2)} +H_R, \nonumber \\
& &H_{44}(t)=E_1(t)+E_2(t)+J(t)-\hbar \Gamma^{(1)} - \hbar \Gamma^{(2)} +H_R,  \nonumber \\
& &H_{23}(t)=H_{32}(t)=2J, \nonumber \\
& &E_i(t)\equiv\hbar \tilde B_z^{(i)}(t), \quad i=1,2. \label{H_matrix}
\end{eqnarray}
$\Gamma^{(1)}$, $\Gamma^{(2)}$, and $H_R$ are operators defined over the boson Hilbert space.\par
In the latter part of this paper, we will use the density matrix formalism of quantum mechanics. Density operators corresponding to the double--qubit system will be denoted by $\rho_S$, whereas density matrices $\rho$ without $\mbox{subscript}\;S$ refer to the composite system (or subspaces thereof) \textit{including} the bosonic environment.

\subsection{Equations of motion}\label{sec:theory-eom}
Since 
\begin{equation}
\left[ H_a(t),H_b(t')\right] =0,
\end{equation}
the propagator for the composite system may be written
in the form 
\begin{eqnarray}
U(t,0)=U_a(t)U_b(t),
\end{eqnarray}
with 
\begin{equation}
U_j(t)=Te^{-\frac{i}{\hbar}\int\limits_0^t {H_j(t')dt'}},\;\mbox{for}\quad j=a,b \quad, \label{U_o}
\end{equation}
and $T$ representing the time--ordering operator.
It is instructive to define two special density matrices corresponding to the composite system for $t=0$,
\begin{eqnarray}
\rho^{(a)}(0) &\equiv& \begin{bmatrix}
\rho_{11}(0) & 0 & 0 & \rho_{14}(0) \\
0 & 0 & 0 & 0 \\
0 & 0 & 0 & 0 \\
\rho_{14}^*(0) & 0 & 0 & \rho_{44}(0) 
\end{bmatrix},\label{rho_o} \\
\rho^{(b)}(0) &\equiv& \begin{bmatrix}
0 & 0 & 0 & 0 \\
0 & \rho_{22}(0) & \rho_{23}(0) & 0 \\
0 & \rho_{23}^*(0) & \rho_{33}(0) & 0 \\
0 & 0 & 0 & 0 
\end{bmatrix}, \label{rho_i}
\end{eqnarray}
similar to $H_a(t)$ and $H_b(t)$.
One can calculate their time evolution as 
\begin{eqnarray}
\rho^{(a)}(t)=U_a(t) \rho^{(a)}(0) U_a^\dag(t), \\
\rho^{(b)}(t)=U_b(t) \rho^{(b)}(0) U_b^\dag(t), 
\end{eqnarray}
by using
\begin{equation}
\left[ H_b(t),\rho^{(a)}(0)\right] = \left[ H_a(t),\rho^{(b)}(0)\right] = 0.
\end{equation}
Thus, if we start with a density matrix of the form of Eq.~\eqref{rho_o} or \eqref{rho_i} the system will remain in the corresponding subspace and we are effectively dealing with two noninteracting two--level systems. By applying the basis transformation given in Table~\ref{tab_mapping}, we are able to map the Hamiltonians $H_a(t)$ and $H_b(t)$ onto
\begin{eqnarray}
H_{a}^{red}(t)&=&-\left( E_1(t) + E_2(t)\right)  \sigma_z + J(t)\openone  \label{H_TLS_o} \\
&+&\sigma_z \left( \hbar\Gamma^{(1)}+\hbar\Gamma^{(2)} \right)+H_R,  \nonumber \\
H_{b}^{red}(t)&=&2J(t) \sigma_z -J(t)\openone \label{H_TLS_i} \\
&+&\sigma_x \left( \hbar\Gamma^{(1)}-\hbar\Gamma^{(2)} + E_2(t) - E_1(t)\right)+H_R.  \nonumber
\end{eqnarray}
It can be seen that for $\Gamma^{(1)}=\Gamma^{(2)}$, {\it i.e.}, for one and the same bath for both spins, the bath terms in Eq.~\eqref{H_TLS_i} cancel. Therefore, the ``interaction part'' of the Hamiltonian $H_b^{red}$ vanishes and no entanglement with the environment occurs. So, for the case of collective decoherence, density matrices of the form
\begin{eqnarray}
\rho_S &=& \begin{bmatrix}
\rho_{S,11} & 0 & 0 & 0 \\
0 & \rho_{S,22} & \rho_{S,23} & 0 \\
0 & \rho_{S,23}^* & \rho_{S,33} & 0 \\
0 & 0 & 0 & \rho_{S,44}
\end{bmatrix} \\
\left( \rho_S \right.&=& \left.\operatorname{tr_R}\left\lbrace \rho \right\rbrace\right)  \nonumber
\end{eqnarray}
are decoherence--free~\cite{Lid98} even if we are dealing with a full Heisenberg--like qubit--qubit coupling (see Eq.~\eqref{H_S_12}). To be precise, the density matrix elements $\rho_{S,11}$ and $\rho_{S,44}$ should be called stationary. This can be seen if we set $\rho_{14}(0)=\rho_{41}(0)=0$ in Eq.~\eqref{rho_o}. Because the propagator $U_a$ is diagonal, it now follows that
\begin{eqnarray}
\rho_{S,11}(t)&=&\rho_{S,11}(0), \nonumber \\
\rho_{S,44}(t)&=&\rho_{S,44}(0). \nonumber
\end{eqnarray}
Furthermore, control in the $\mbox{$\rho^{(a)}$ subspace}$ via the qubit--qubit coupling is not possible because there is only a trivial ({\it i.e.}, proportional to $\openone$) dependence on $J(t)$ in Eq.~\eqref{H_TLS_o}.
\begin{table}[H]
\begin{center}
\begin{tabular}{lccccc}
\hline
\hline
\multicolumn{4}{l}{\parbox[l]{4.4cm}{Four--level system}} & \parbox[c]{2.4cm}{Reduced TLS}  & Subspace\\
\hline
$|\psi^+\rangle$ & $\doteq$ & $\frac{1}{\sqrt{2}}\left( 0,1,1,0\right)^T$ &$\rightarrow$& $\left( 1,0\right) ^T$ & \\
$|\psi^-\rangle$ & $\doteq$ & $\frac{1}{\sqrt{2}}\left( 0,1,-1,0\right)^T$ &$\rightarrow$& $\left( 0,1\right) ^T$ & \raisebox{2ex}[-2ex]{$\rho^{(b)}$}\vspace{1mm}\\
\hline
$|11\rangle$ & $\doteq$ & $\left( 1,0,0,0\right)^T$ &$\rightarrow$& $\left( 1,0\right) ^T$ & \\
$|00\rangle$ & $\doteq$ & $\left( 0,0,0,1\right)^T$ &$\rightarrow$& $\left( 0,1\right) ^T$ & \raisebox{2ex}[-2ex]{$\rho^{(a)}$}\\
\hline
\hline
\end{tabular}
\caption{\label{tab_mapping} Mapping of the basis for the full four--level system to one for the reduced TLSs.}
\end{center}
\end{table}
\noindent
If the initial state of the double dot is any valid superposition of $\rho^{(a)}(0)$ and $\rho^{(b)}(0)$,
it is possible to formulate and solve the master equations for each of the two subsystems independently. However, in the following, we are going to solve the full four--level dynamics.\par 

In the interaction picture (operators denoted by a tilde), using
\begin{eqnarray}
U_{o}(t,0)&=&\operatorname{T}e^{-\frac{i}{\hbar}\int\limits_0^t {\left(H_S(t')+H_R\right)dt'}}, \label{prop_ges}\\
\tilde \rho(t)&=&U_o^\dag(t,0) \rho(t) U_o(t,0) \label{def_rho_I},\\
\tilde H_{SR}(t)&=&U_o^\dag(t,0) H_{SR}(t) U_o(t,0) \label{H_SR_I}, 
\end{eqnarray}
the master equation in the Born-Markov approximation for the present system-bath interaction is of the form~\cite{Car02}
\begin{eqnarray}
& &\frac{d}{dt} \tilde \rho_S(t)= \label{master_born_markov} \\
& &-\frac{1}{\hbar^2} \int_0^t{dt'\operatorname{tr_R}\left\lbrace \left[\tilde H_{SR}(t),\left[\tilde H_{SR}(t'),\tilde \rho_S(t) \otimes \tilde \rho_R(0) \right] \right] \right\rbrace  }. \nonumber
\end{eqnarray}
When evaluating the double commutator one encounters the correlation functions
\begin{equation}
\left\langle {\tilde \Gamma^{(i)}(t) \tilde \Gamma^{(i)}(t') } \right\rangle _R=\operatorname{tr_R}\left\lbrace \tilde \Gamma^{(i)}(t) \tilde \Gamma^{(i)}(t') \tilde \rho_R(0) \right\rbrace. \label{corrf0}
\end{equation}
These can be calculated analytically if we use an Ohmic spectral density,~\cite{Leg87,Weiss99}
\begin{eqnarray}
\sum_k {\left\lbrace g_k^2 ...\right\rbrace} &\rightarrow& \int\limits_0^\infty{d\omega \left\lbrace J\left( \omega \right) ...\right\rbrace}, \nonumber \\
J(\omega) &\rightarrow & \eta \omega e^{-\frac{\omega}{\omega_c}}. \label{ohmic}
\end{eqnarray}
$\omega_c$ is a cutoff frequency and $\eta$ is a parameter which describes the coupling strength of the bosons to the qubit. The specific value for $\omega_c$ depends on the physical nature of the dephasing mechanism.\par
Up to second order in the system--reservoir interaction, we may use the equilibrium form for $\rho_R$ in the calculation of the correlation functions,
\begin{eqnarray}
\rho_R^{(i)}&=&\prod_k {\left[{\left( 1-\exp\left({-\hbar \omega_k^{(i)} \beta}\right)\right) \exp\left(-\hbar \omega_k^{(i)} \beta b_k^{(i)\dag} b_k^{(i)\vphantom{\dag}} \right)   }\right] }, \label{rho_R} \nonumber \\
\rho_R&=&\rho_R^{(1)} \otimes \rho_R^{(2)}. 
\end{eqnarray}
With $\beta=\frac{1}{k_B T}$, we get
\begin{eqnarray}
& &\left\langle {\tilde \Gamma^{(i)}(t) \tilde \Gamma^{(i)}(t') } \right\rangle _R=\frac{2\eta}{\hbar^2 \pi \beta^2}\left\lbrace \psi' \left( 1+ \frac{1-i \omega_c\left( t-t'\right) }{\hbar \omega_c \beta}\right) \right.\nonumber \\
& &+\left.\psi' \left( \frac{1+i \omega_c\left( t-t'\right) }{\hbar \omega_c \beta}\right)\right\rbrace, \label{corrf1}
\end{eqnarray}
where $\psi'$ is the derivative of the digamma function.~\cite{wolfram} The calculation of Eq.~\eqref{corrf1} is outlined in Appendix~\ref{app:corrf}.
\par 
Depending on the physical nature of the interaction, one may use more sophisticated formulations of the spectral density.~\cite{Brand02} In Ref.~\onlinecite{West04}, the dissipative interaction between the orbital degrees of freedom and an acoustic phonon bath was investigated. This interaction leads to an effective spin--phonon interaction which is related to a more complicated spectral density. However, using these results would give correlation functions which cannot be calculated analytically.\par
In the following calculations we will choose $T=50$~mK for the temperature. At this temperature, the spin--flip energy $\Delta E_Z$ (at $B=1$~T) is approximately five times larger than $k_B T$. So, we can neglect spin flips and concentrate on dephasing processes. Dephasing is an interesting phenomenon because decoherence associated with this process can occur even when there is no real energy exchange with the environment. Also, quantum vacuum fluctuations can cause dephasing.~\cite{Pal96,Buett01,Rei02}

\subsection{Optimization strategy}\label{sec:theory-optstrategy}
Assume first that we want to trap the system in the Bell state $|\psi^+\rangle$, where
\begin{equation}
|\psi^\pm\rangle = \frac{1}{\sqrt{2}} (|1\rangle \otimes |0\rangle \pm |0\rangle \otimes |1\rangle) \doteq \frac{1}{\sqrt{2}}\left( 0,1,\pm 1,0\right)^T.
\end{equation}
Based on the discussion above, the strategy is to make this state to the nondegenerate ground state of the isolated driven (\textit{i.e.}, decoupled from the bath) spin system.~\cite{Poetz1} From Eq.~\eqref{H_TLS_i} and Table~\ref{tab_mapping}, we see that for $J(t)<0$, the state corresponding to $|\psi^+\rangle$ indeed becomes the ground state of the nondissipative reduced $\mbox{TLS $b$}$. Thus, if we set $J(t)$ large and negative, we would infer that this state is rather robust with respect to dephasing because the levelsplitting between $|\psi^+\rangle$ and $|\psi^-\rangle$ is $4|J(t)|$. (We set $E_1(t)=E_2(t)$ for simplicity.)\par
This statement is readily verified numerically. Suppose that we start with the 2x2 density matrix for TLS $b$ (see Sec.~\ref{sec:theory-eom}),
\begin{equation}
\rho^{(b)}_S(0)=\begin{bmatrix}
1 & 0 \\
0 & 0
\end{bmatrix} \leftrightarrow |\psi^+\rangle \langle \psi^+|. \label{rho_i_initial}
\end{equation}
Without the qubit--qubit interaction ({\it i.e.}, $J=0$), it evolves into the mixture
\begin{equation}
\rho^{(b)}_S(\infty)=\begin{bmatrix}
\frac{1}{2} & 0 \\
0 & \frac{1}{2}
\end{bmatrix} \leftrightarrow \frac{1}{2} |\psi^+\rangle \langle \psi^+|+\frac{1}{2} |\psi^-\rangle \langle \psi^-|.
\end{equation}
\par
\begin{center}
\begin{figure}[h]
\begin{tabular*}{8.5cm}{ll}
(a) & \\
 & \includegraphics[width=8cm]{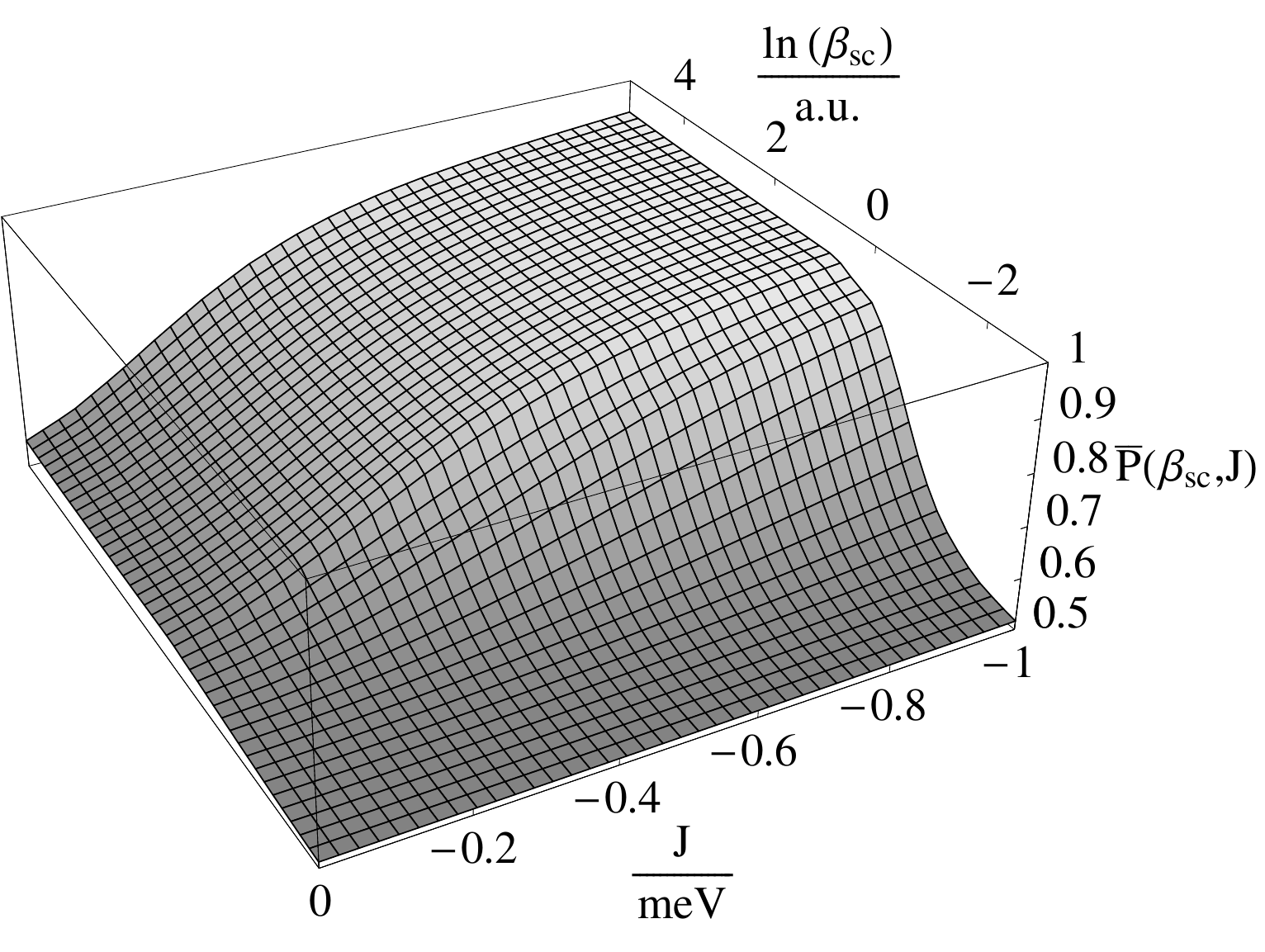} \\
(b) & \\
 & \includegraphics[width=8cm]{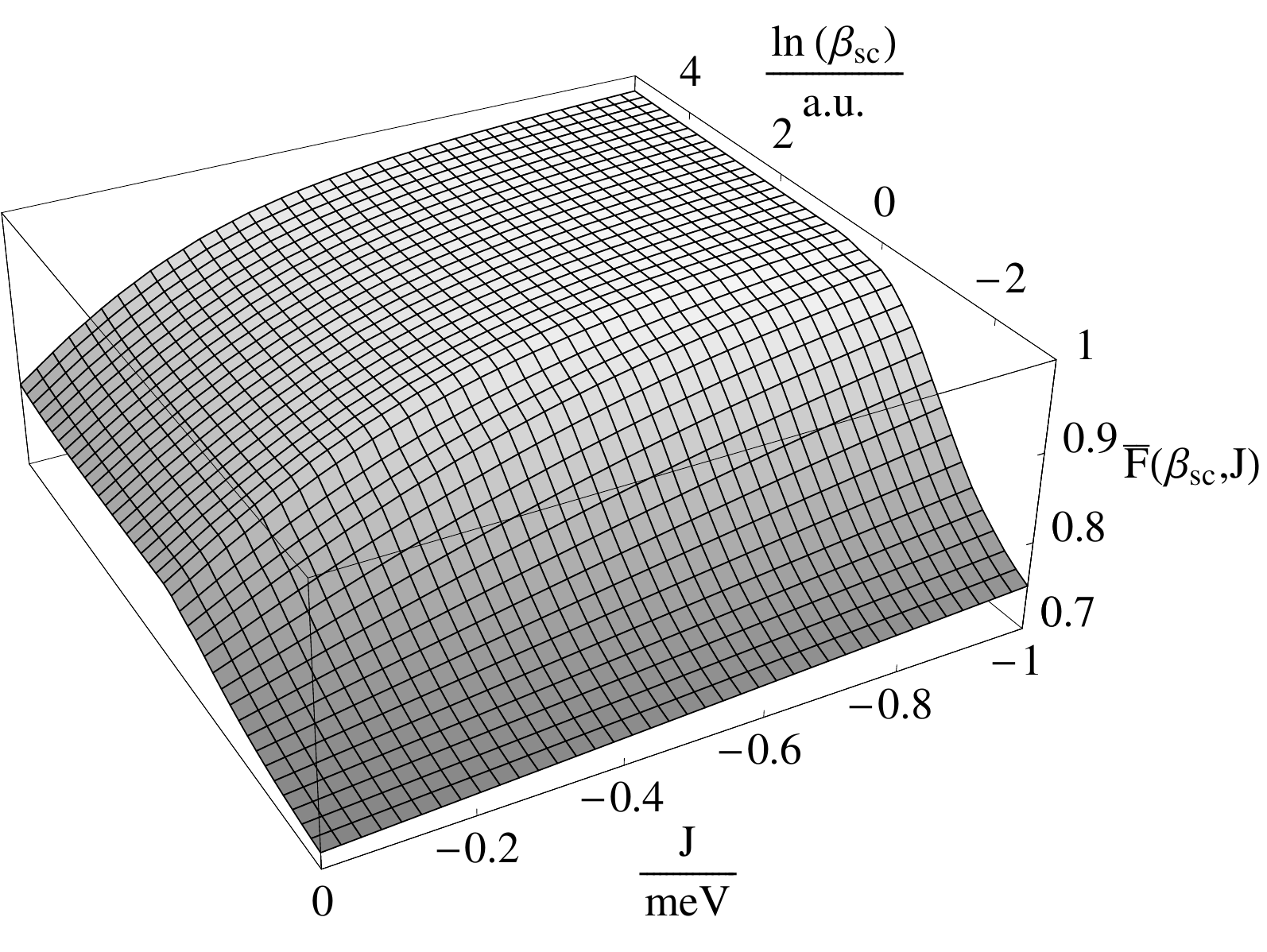} 
\end{tabular*}
\caption{\label{PJbeta}(a) Time--averaged purity $\bar P$ and (b) fidelity $\bar F$ vs qubit--qubit coupling $J$ and the logarithm of inverse temperature. $\eta=0.1$, $t_f=\frac{2}{\omega_{sc}}$ and $\omega_{c}=20 \einheit{meV}$.}
\end{figure}
\end{center}

 Now, let us set $J(t)=J=const$ and examine the dependence of the time--averaged purity,
\begin{equation}
\bar P\left( t_f \right) = \frac{1}{t_f} \int\limits_0^{t_f} {\operatorname{tr_S} \left\lbrace \rho^{(b)}_s(t)^2\right\rbrace dt},
\end{equation}
and the time--averaged fidelity,~\cite{Niel02}
\begin{equation}
\bar F\left(t_f \right) = \frac{1}{t_f} \int\limits_0^{t_f} {\operatorname{tr_S} \left\lbrace \sqrt{\left( \rho^{(b)}_S(0)\right)^{\frac{1}{2}}\rho^{(b)}_S(t)\left(\rho^{(b)}_S(0)\right) ^{\frac{1}{2}}}\right\rbrace dt},
\end{equation}
on the qubit--qubit coupling $J$ and the inverse temperature $\beta_{sc}=\frac{\hbar \omega_{sc}}{k_B T}$ ($\hbar \omega_{sc}=1\einheit{meV}$) with $\rho^{(b)}_S(0)$ given by Eq.~\eqref{rho_i_initial}. (In the following, physical quantities with a subscript $sc$ are scaled with respect to $\omega_{sc}$.)  The result can be seen in Fig.~\ref{PJbeta}, keeping in mind that for bipartite systems,  
\begin{eqnarray}
\bar P(t_f)_{max}&=&1, \nonumber \\
\bar P(t_f)_{min}&=&\frac{1}{2}.
\end{eqnarray}
For ``low'' temperature (up to $T \approx 15 \einheit{K}$), good results can be obtained by setting $J \approx -1 \einheit{meV}$ (qubit--qubit couplings of several $100 \einheit{$\mu$eV}$ have been reported in Ref.~\onlinecite{Elz}). However, it should be noted that for $T\approx15 \einheit{K}$ and $B_z=1 \einheit{T}$, spin--flip processes constitute the main mechanism of decoherence and our model accounting for pure dephasing only is no longer valid.
\par
The conclusion of Secs.~\ref{sec:theory-eom} and \ref{sec:theory-optstrategy} may be summarized as follows. Given a Hamiltonian of the form of Eq.~\eqref{H_matrix}, suppression of pure dephasing of the maximally entangled states $|\psi^\pm\rangle$ is possible by tuning of the qubit--qubit interaction. Explicit results of this strategy for state trapping and steering will be given in the next section.\par
On the other hand, dephasing of states,
\begin{equation}
|\Phi^\pm\rangle = \frac{1}{\sqrt{2}} (|1\rangle \otimes |1\rangle \pm |0\rangle \otimes |0\rangle) \doteq \frac{1}{\sqrt{2}}\left( 1,0,0,\pm 1\right)^T,
\end{equation}
cannot be reduced by means of changing $J$. To do so, the qubit--qubit Hamiltonian has to display anisotropy, e.g.,
\begin{equation}
{H_S}^{(12)}=J(t) \sigma_x^{(1)}\otimes\sigma_x^{(2)}.
\end{equation}
With the isotropic Hamiltonian [Eq.~\eqref{H_matrix}], it is also impossible to prepare dephasing--insensitive entangled states of the form
\begin{equation}
\frac{1}{\sqrt{2}} (|1\rangle \otimes |0\rangle \pm i|0\rangle \otimes |1\rangle) \doteq \frac{1}{\sqrt{2}}\left( 0,1,\pm i,0\right)^T.
\end{equation}
For this task, a coupling,
\begin{equation}
{H_S}^{(12)}=J(t) \sigma_x^{(1)}\otimes\sigma_y^{(2)},
\end{equation}
would be needed. Hence, to trap an arbitrary Bell state by this strategy, a detailed control of the spin--spin coupling is required.  Realization feasibility depends on the physical nature of the double qubit.
\section{Numerical results}\label{sec:results}

We now investigate feasibility of the strategy discussed above at the example of state trapping and steering into the Bell state $|\psi^+\rangle$ by solving Eq.~\eqref{master_born_markov} numerically.  In order to quantify the degree of success, we use several well--established quantities. A convenient measure for the degree of entanglement of bipartite systems is the concurrence.~\cite{Wo98}
The concurrence $C$ equals 1 if the qubits are maximally entangled. It becomes 0 if the system factorizes or, for the case of a mixed density matrix, if the system can be represented by a mixture of factorizable pure states. Another important measure is the purity $P$, which gives information whether the system is in a pure state or in a mixture.~\cite{Schwabl} A purity of 1 indicates a pure state, while $\frac{1}{2} \leq P<1$ occurs if the system is in a mixed state.
\subsection{Trapping}\label{sec:results-trapping}
\begin{figure}[h]
\begin{tabular*}{8.5cm}{ll}
(a) & (b) \\
\includegraphics[width=4.25cm]{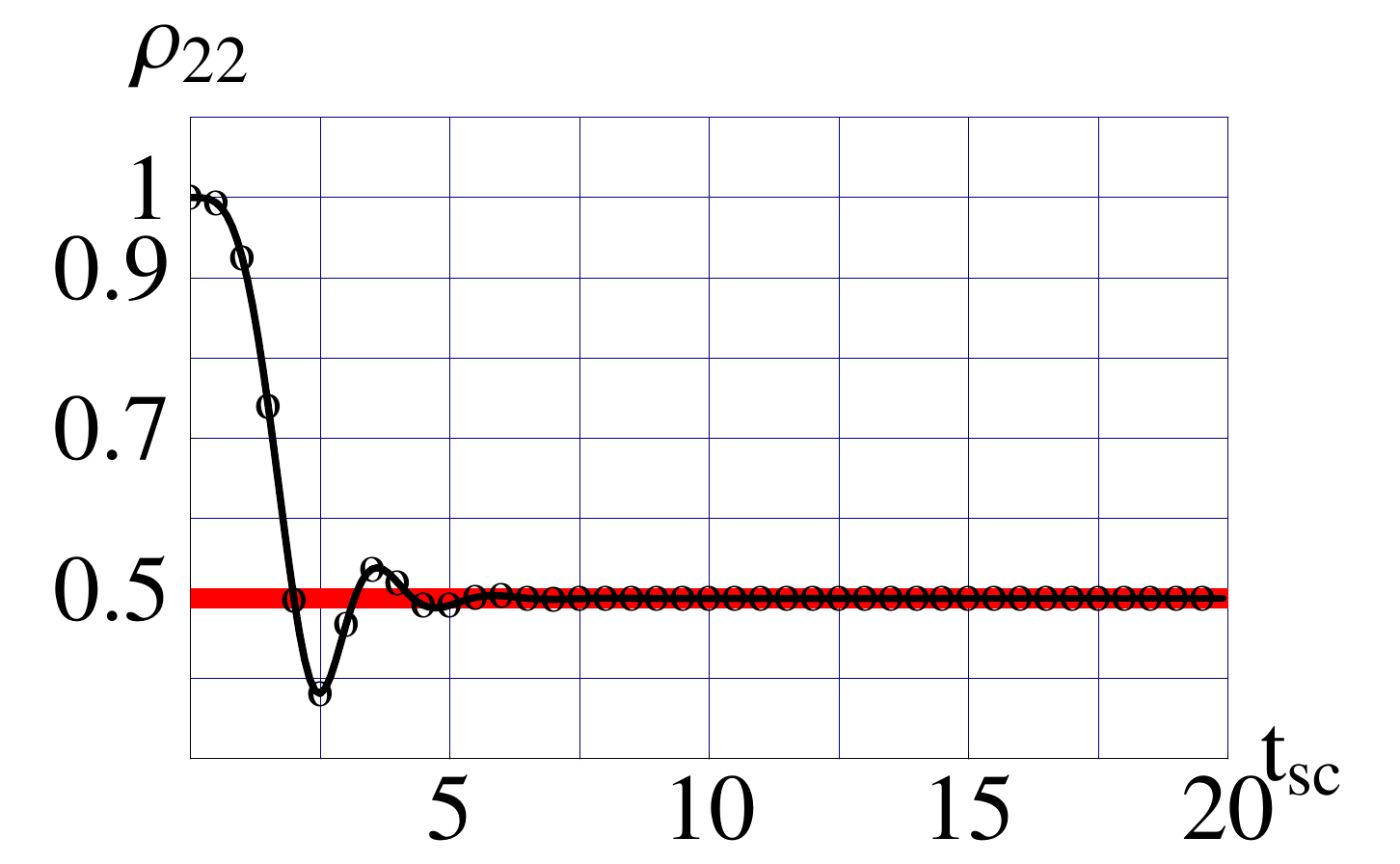} & \includegraphics[width=4.25cm]{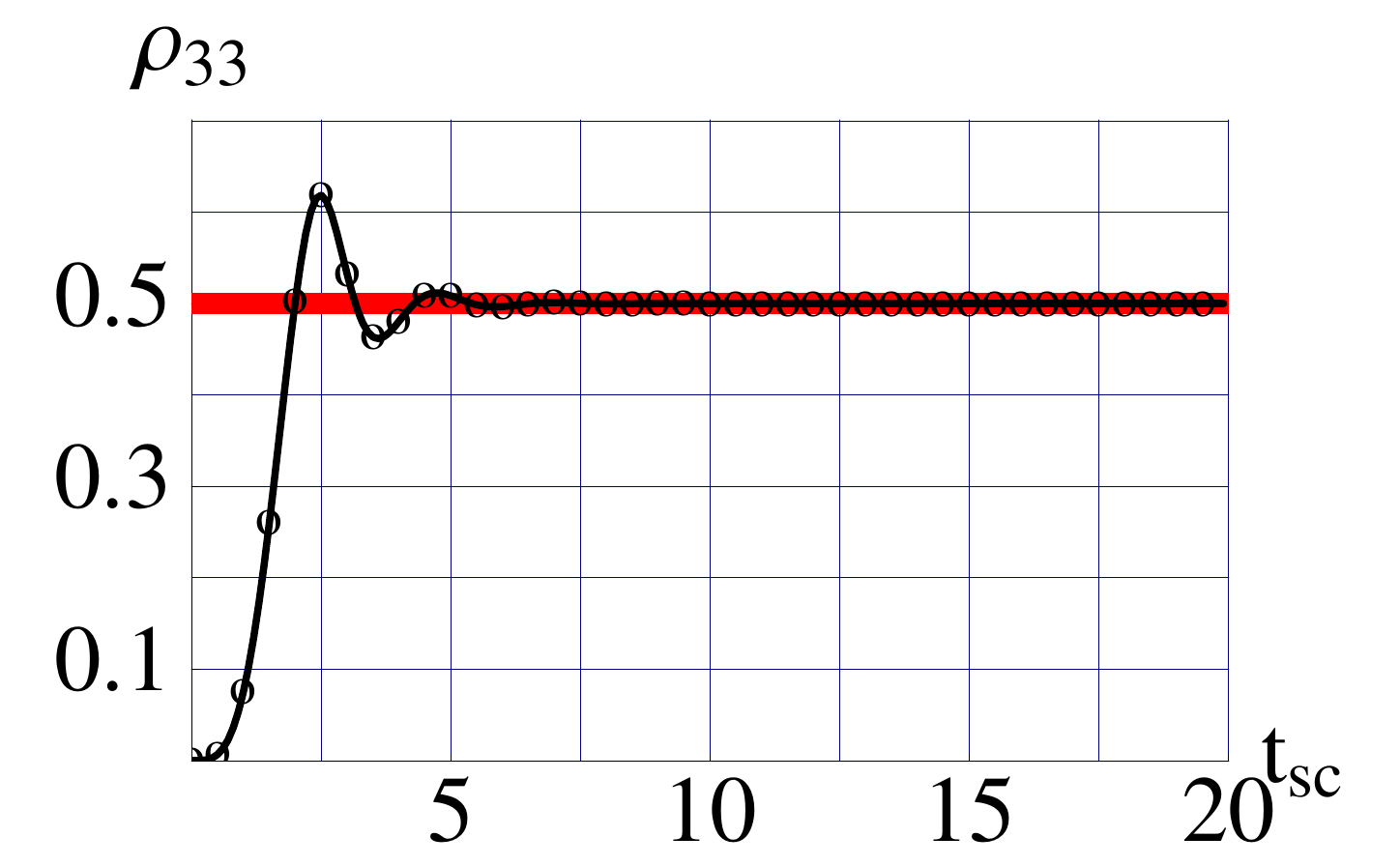} \\
(c) & (d) \\
\includegraphics[width=4.25cm]{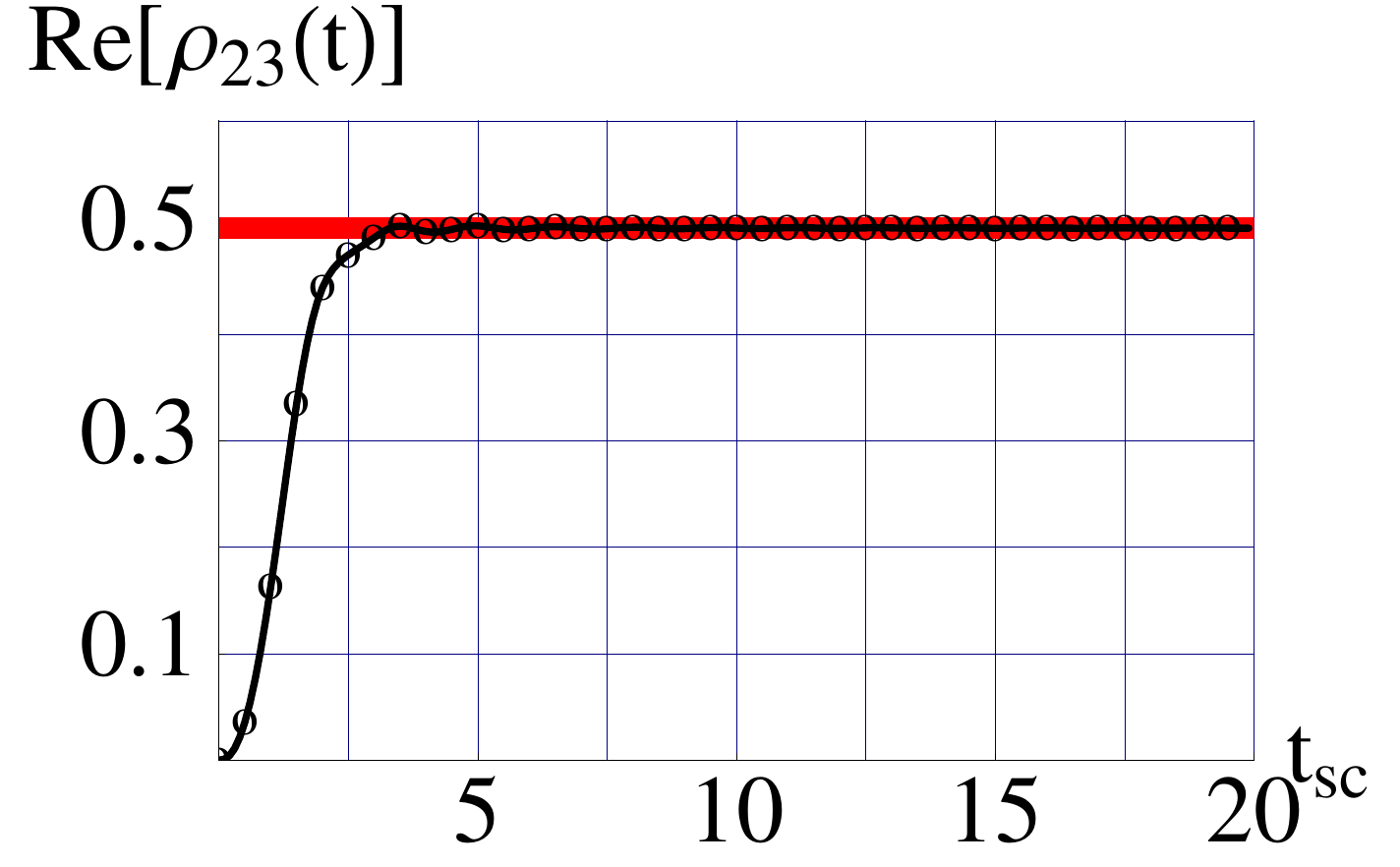} & \includegraphics[width=4.25cm]{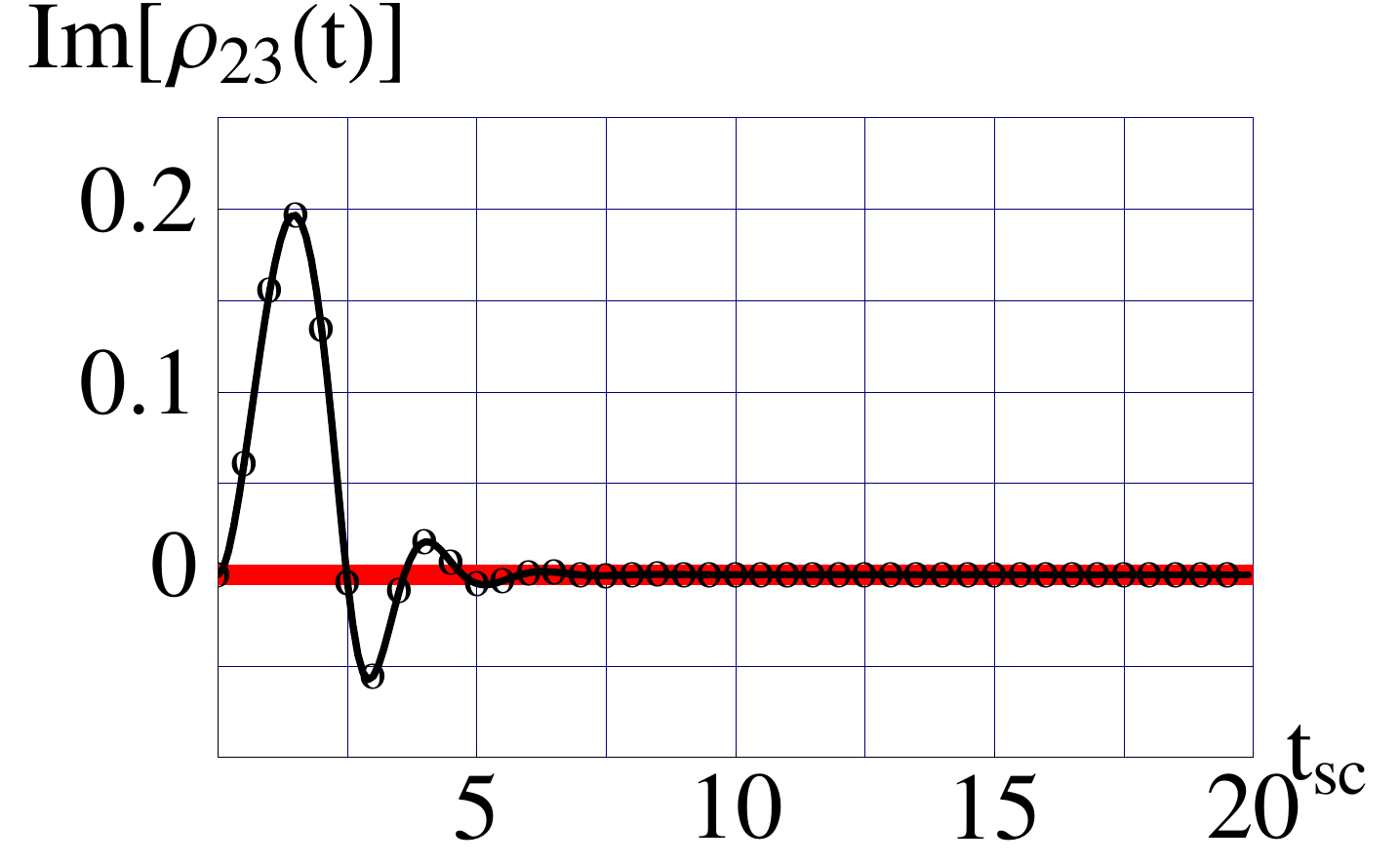}\\
(e) & (f) \\
\includegraphics[width=4.25cm]{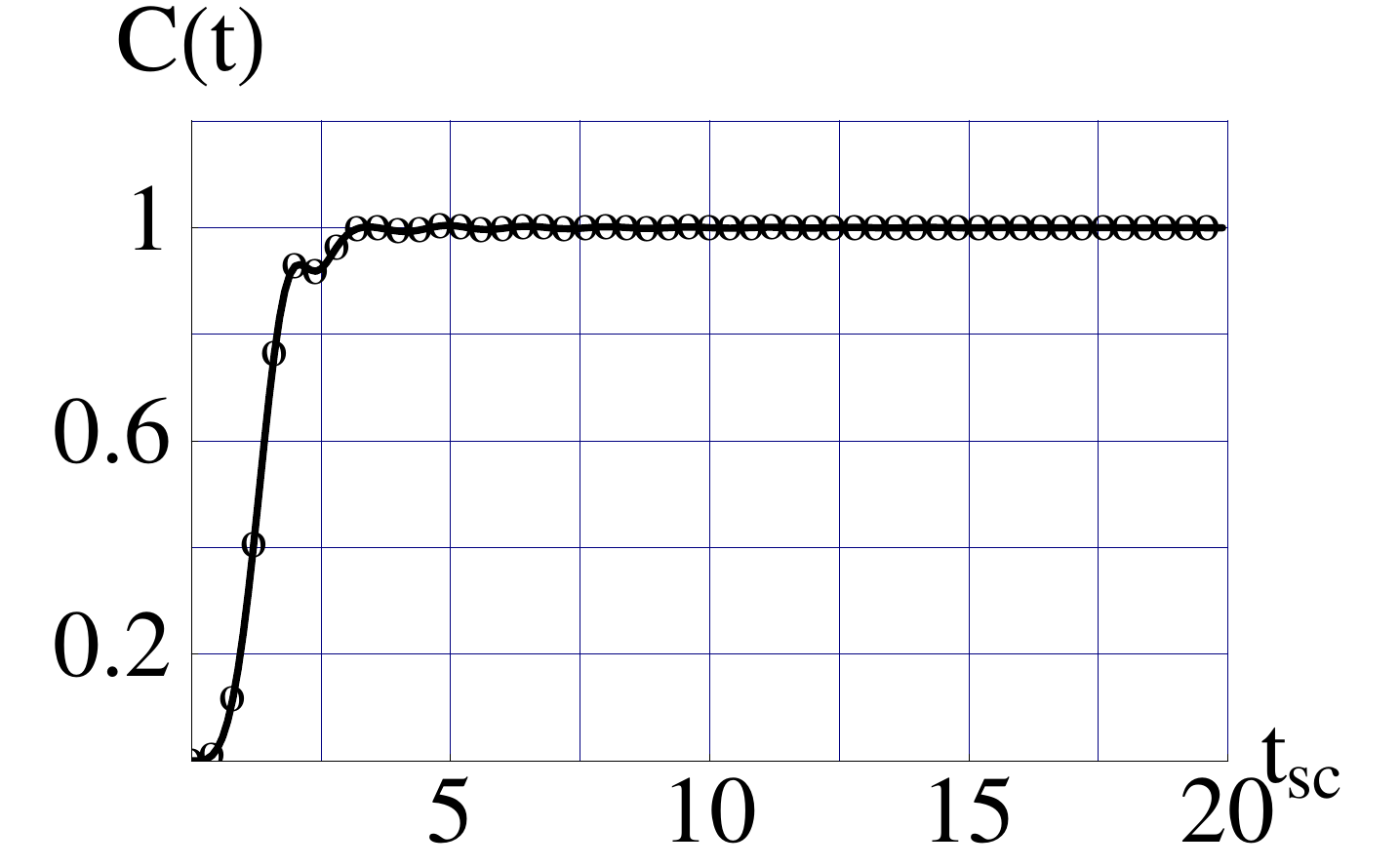} & \includegraphics[width=4.25cm]{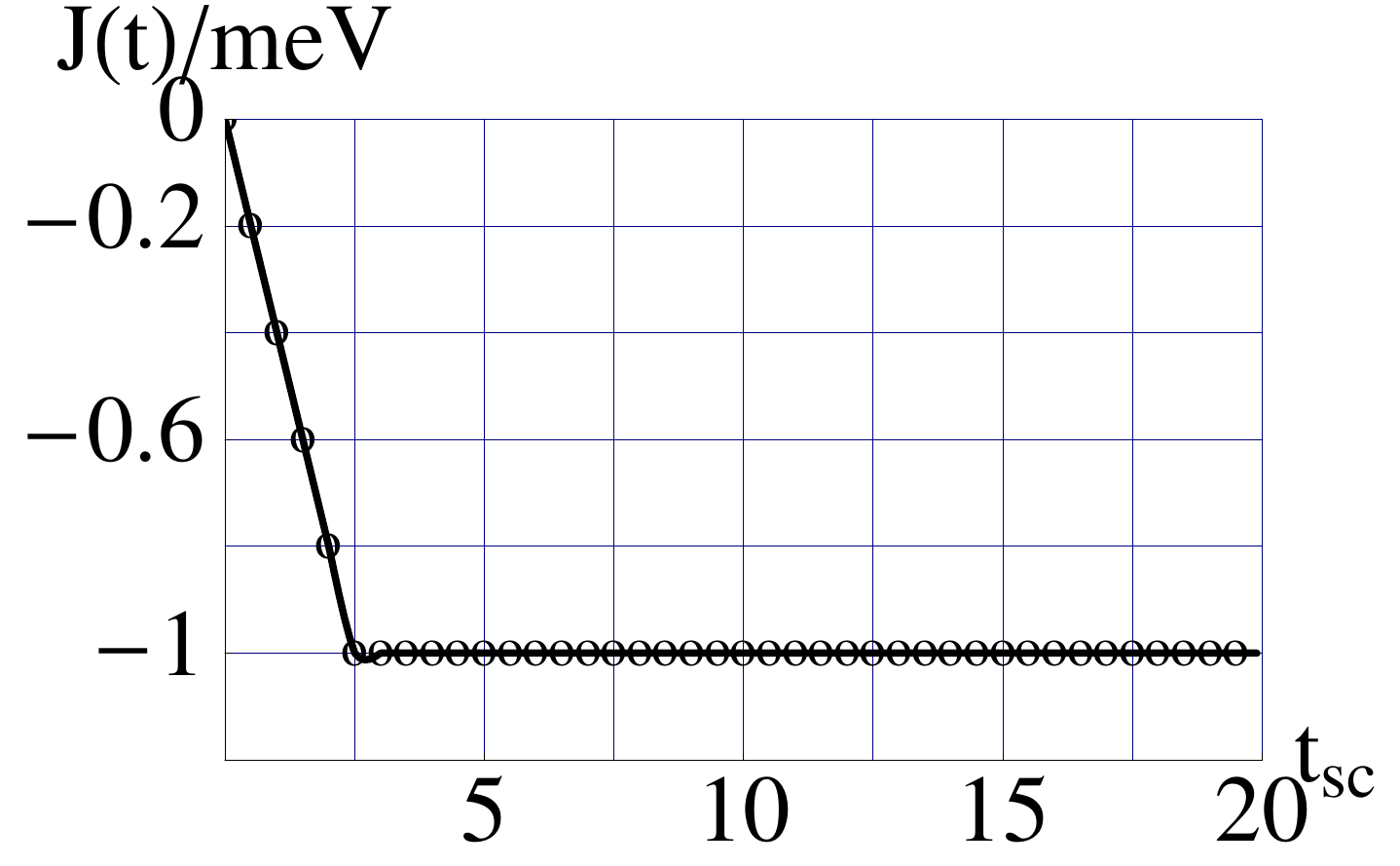}
\end{tabular*}
\caption{\label{fig:trapping}Red (thick solid) lines indicate the desired values of the corresponding density matrix elements. (a)~The trajectory of the diagonal element $\rho_{22}$, (b)~the diagonal density matrix element $\rho_{33}=1-\rho_{22}$, (c) the real part of the off--diagonal density matrix element $\operatorname{Re}\left[ \rho_{23}\right]=\operatorname{Re}\left[ \rho_{32}\right]$, and (d) the corresponding imaginary part $\operatorname{Im}\left[ \rho_{23}\right]=-\operatorname{Im}\left[ \rho_{32}\right]$. (e)~The concurrence vs time and (f)~the control field $J(t)$ vs time.}
\end{figure}

The aim is to steer the double-spin system into the entangled state $|\psi^+\rangle$ and to trap it there starting from the initial state,
\begin{equation}
|\psi_I\rangle \langle\psi_I| \doteq \begin{bmatrix}
0 & 0 & 0 & 0 \\
0 & 1 & 0 & 0 \\
0 & 0 & 0 & 0 \\
0 & 0 & 0 & 0 \\
\end{bmatrix} \rightarrow |\psi^+\rangle \langle\psi^+| \doteq \begin{bmatrix}
0 & 0 & 0 & 0 \\
0 & \frac{1}{2} & \frac{1}{2} & 0 \\
0 & \frac{1}{2} & \frac{1}{2} & 0 \\
0 & 0 & 0 & 0 \\
\end{bmatrix}. \label{di_state}
\end{equation}
As outlined in the last section, we use variation of the qubit--qubit coupling $J(t)$ to steer the system into the entangled state $|\psi^+\rangle$ and to trap it by making it insensitive to pure dephasing arising from the interaction [Eq.~\eqref{H_SR}]. We start with a control field $J(0)=0$. The selected time dependence of the control field is shown in Fig.~\ref{fig:trapping}(f).
\begin{table}[h]
\begin{center}
\begin{tabular}{p{0.8cm}p{1cm}cccp{0.8cm}p{0.8cm}}
\hline
\hline
$\eta$		& 	$\hbar \omega_{sc}$		&	T					&	$t_{f}=t_{f,sc} / \omega_{sc}$				&	$\hbar \omega_c$	&	$B^{(1)}_{z}$	&	$B^{(2)}_{z}$ \\
\hline
0.08				&	1meV &	$50 \einheit{mK}$	&	$13.04 \einheit{ps}$&	$20 \einheit{meV}$	&	$1 \einheit{T}$		&	$1 \einheit{T}$ \\
\hline
\hline
\end{tabular}
\caption{\label{tab2} Physical quantities used to calculate the data shown in Figs.~\ref{fig:trapping}(a)--\ref{fig:trapping}(f).}
\end{center}
\end{table}
Figures.~\ref{fig:trapping}(a)-\ref{fig:trapping}(d) show that steering to and trapping in the entangled state $|\psi^+\rangle$ using this technique works very well. Convergence of the off--diagonal density matrix elements is generally better than for the diagonal elements. The concurrence shown in Fig.~\ref{fig:trapping}(e) demonstrates that practically perfect  entanglement ($C\approx 1$) can be achieved.
Furthermore, the achieved entanglement is stable with respect to pure dephasing.\par
\subsection{Optimized driving and trapping}\label{sec:results-trapping+optimization}
\begin{figure}[h]
\begin{tabular*}{8.5cm}{ll}
(a) & (b) \\
\includegraphics[width=4.25cm]{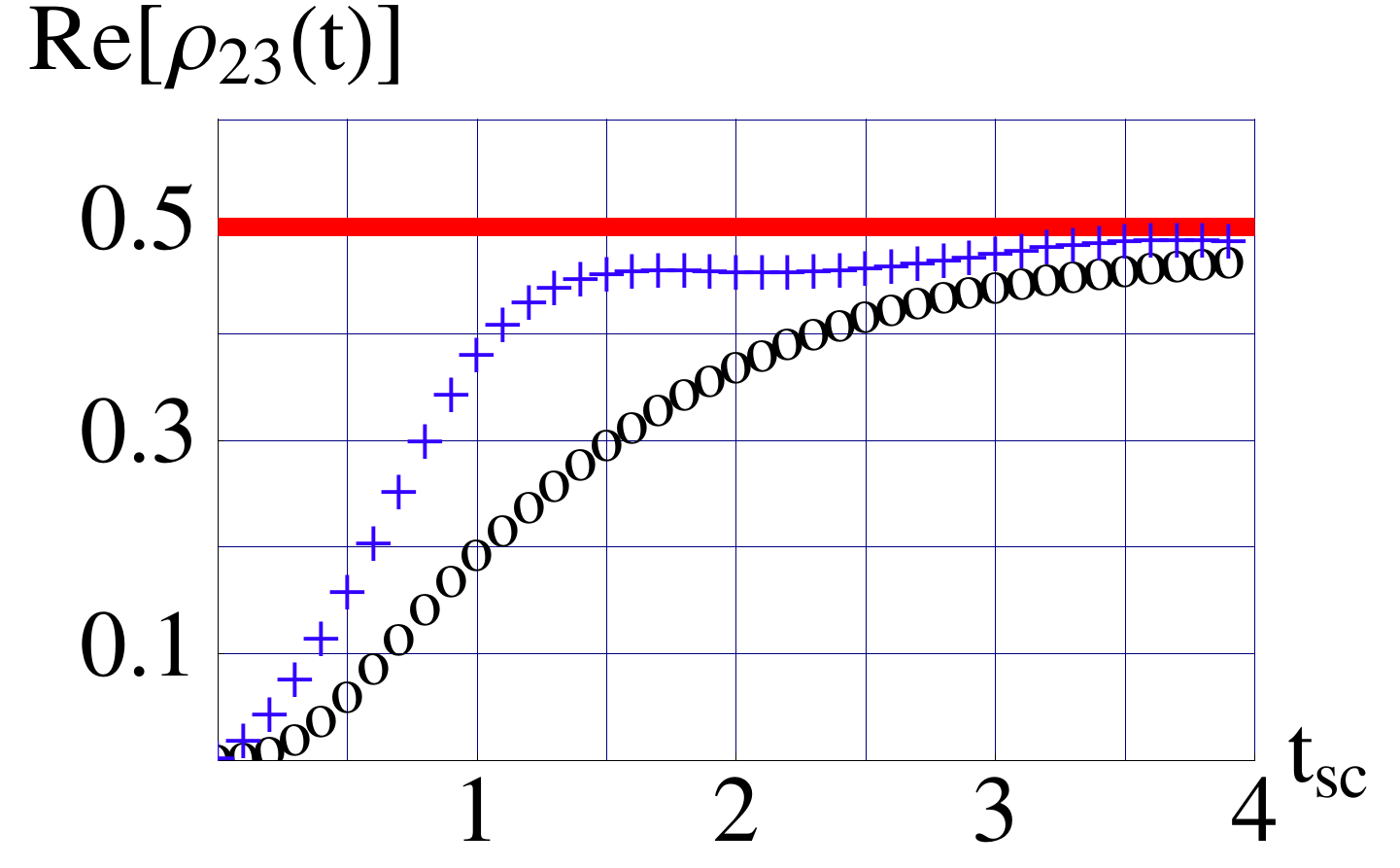} & \includegraphics[width=4.25cm]{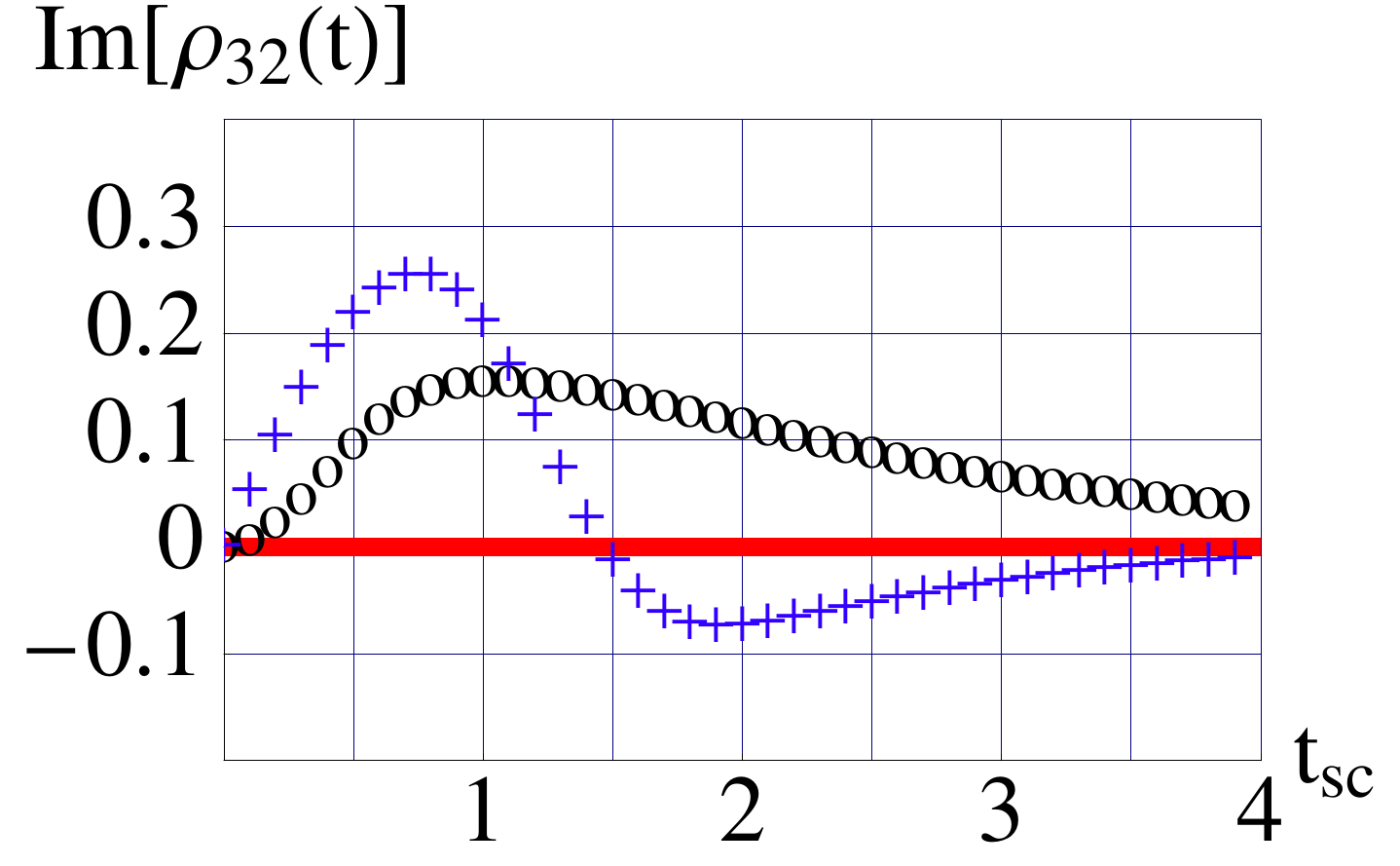}\\
(c) & (d) \\
\includegraphics[width=4.25cm]{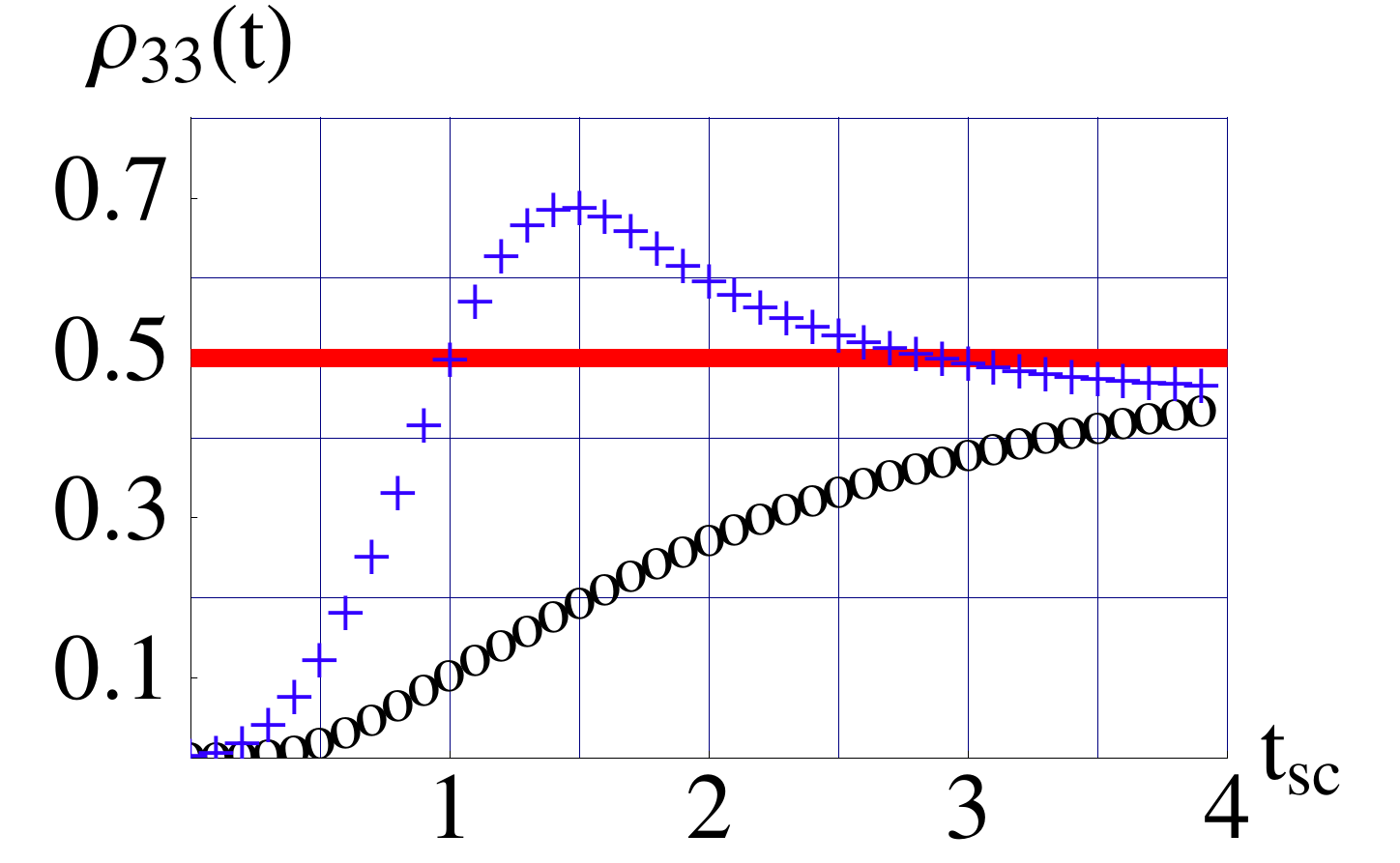} & \includegraphics[width=4.25cm]{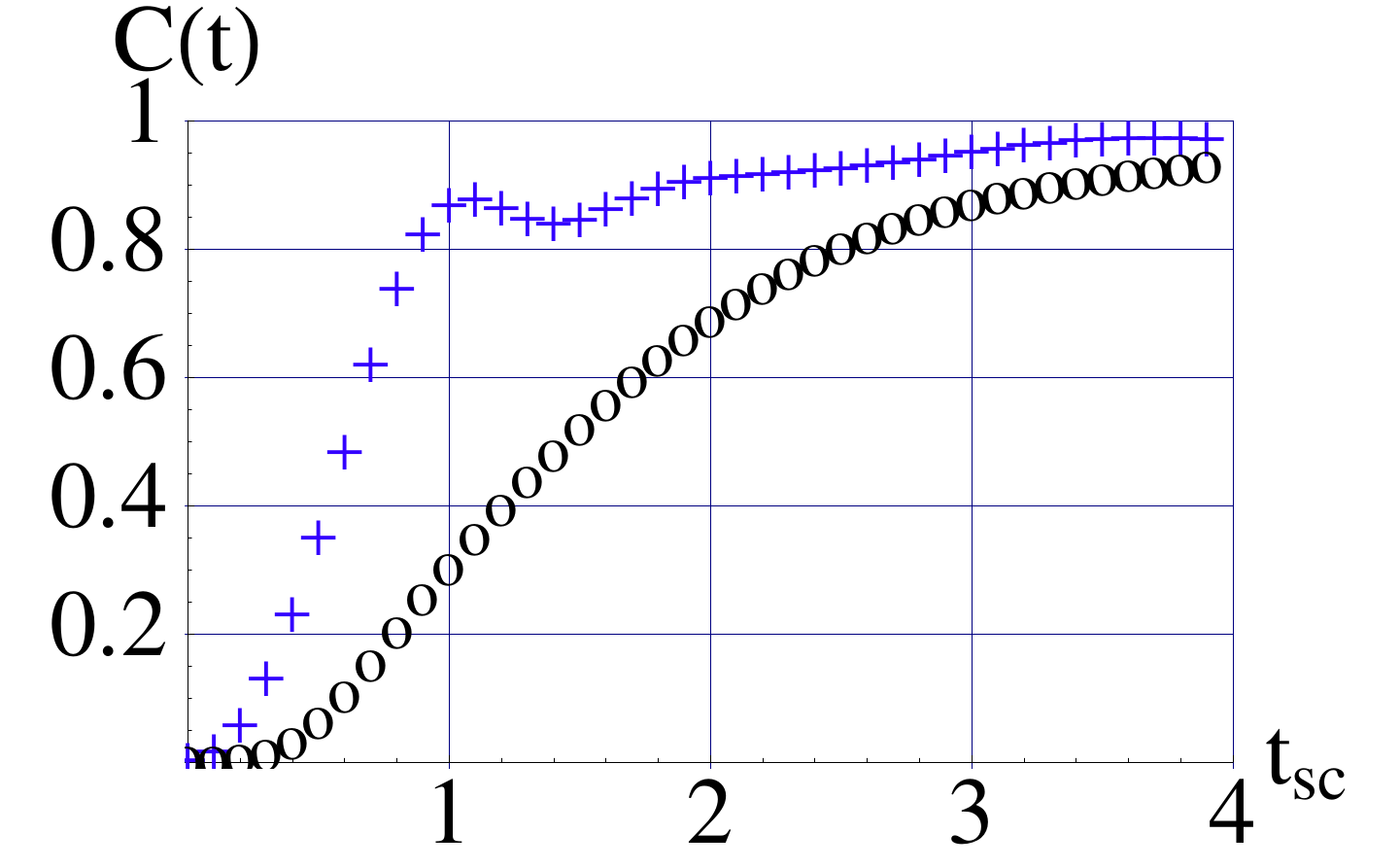}\\
(e) & (f) \\
\includegraphics[width=4.25cm]{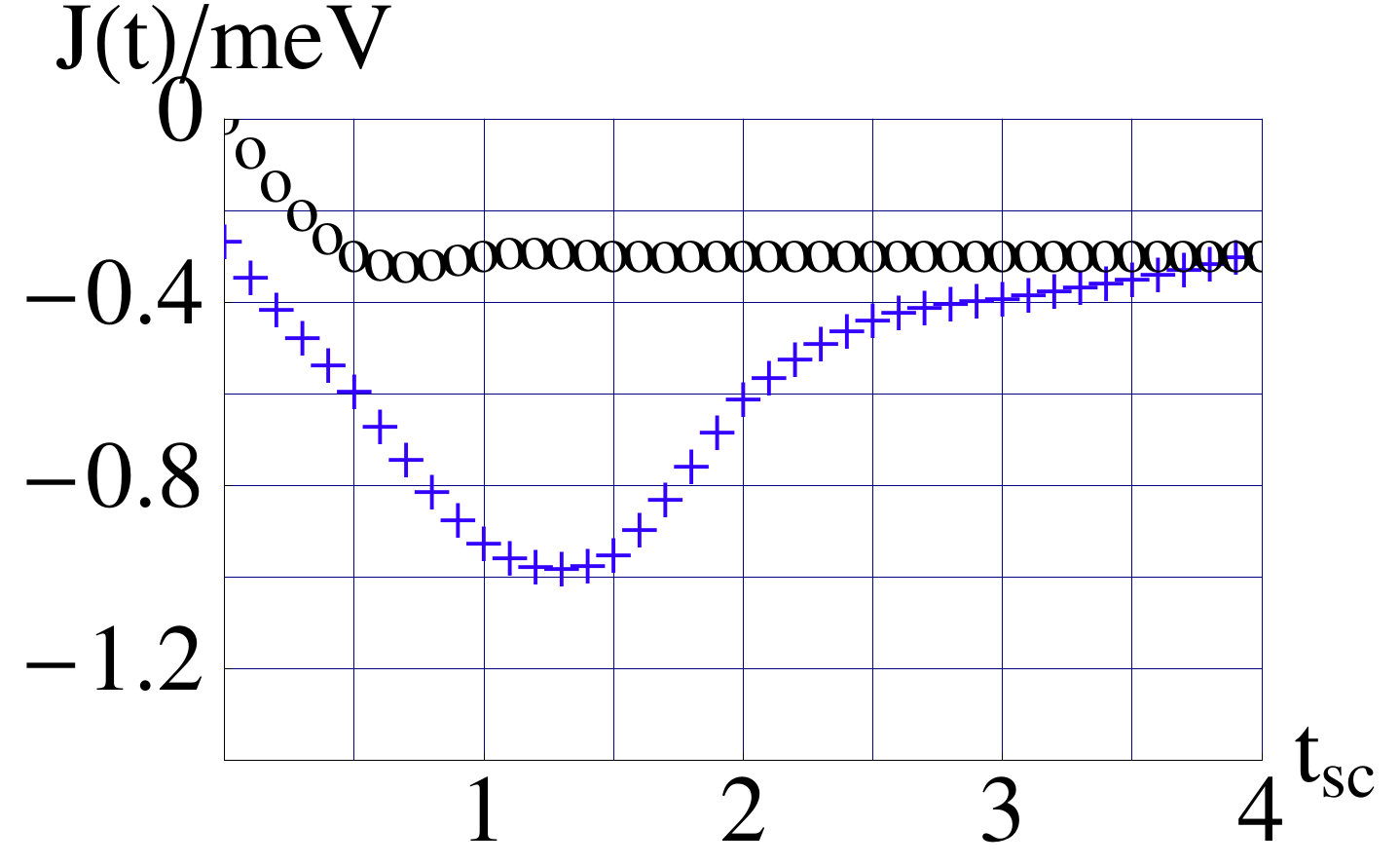} & \includegraphics[width=4.25cm]{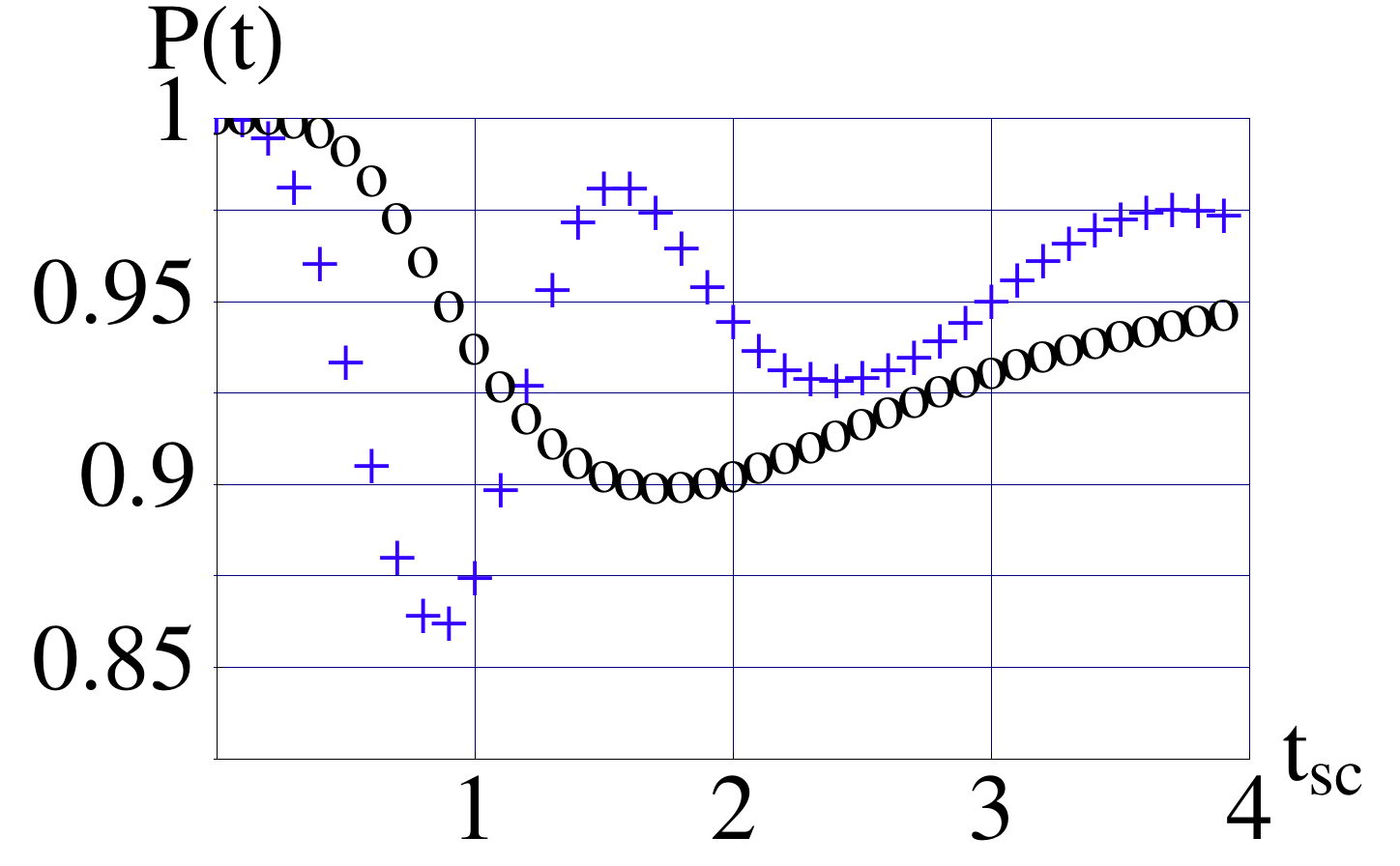}
\end{tabular*}
\caption{\label{fig:trap+opt}Red (thick solid) lines indicate the desired values of the corresponding density matrix elements. (+,~blue) data points denote the optimized case and (o,~black) ones denote the initial data or guess. (a)~The trajectory of the off-diagonal element $\operatorname{Re}\left[ \rho_{23}\right] $, (b)~the imaginary part of $\rho_{32}$, (c)~the diagonal element $\rho_{33}=1-\rho_{22}$, (d)~the concurrence vs time, (e)~the control field $J(t)$, and  (f)~the purity of the system vs time.}
\end{figure}

In this section, we utilize a conjugate gradient (CG) algorithm to optimize the control field plotted in Fig.~\ref{fig:trap+opt}(e).~\cite{NrC} This can be done by formulation of a cost functional~\cite{Poetz3}
\begin{eqnarray}
& &\mathcal{F}\left[ K \right] = \int\limits_{0}^{t_f}{dt \vectornorm {\rho_S(t)-\rho_{S,d}(t)}}^2+\int\limits_{0}^{t_f}{dt \alpha(t) K(t)^2} \nonumber \\
& &\mbox{with}\quad\vectornorm {\vec x(t)}^2=\vec x(t) \cdot \vec x^*(t), \label{J_trapping}
\end{eqnarray}
where $\mathcal{F}$ denotes the cost functional and $K$ the control field. In this notation, the density matrix has been cast into vector form for convenience. Here, we demand that the trajectory of the density matrix follows a given path in time [$\rho_{S,d}(t)$]. The function $\alpha(t)$ is used to avoid an unbounded intensity of the control field during the optimization procedure.\par The problem of optimization can now be stated formally as follows:
\begin{equation}
K^{*}=\mathop{\operatorname{argmin}}\limits_{K}\left( \mathcal{F}\left[ K \right]\right), \label{min_problem}
\end{equation}
where $K^{*}$ is the solution of the minimization problem. In general, it is quite difficult to find a solution to Eq.~\eqref{min_problem} because the cost functional can be very complicated as is indicated by the structure of Eq.~\eqref{J_trapping}. We used a Polak--Ribiere--type conjugate gradient method to search for a minimum of the cost functional.~\cite{NrC}\par
To apply the CG algorithm, the gradient of the cost functional has to be calculated. That can be done, e.g., using a Hamiltonian approach which leads to a set of conjugate differential equations to those for $\rho_S(t)$.~\cite{Poetz2} These have to be solved backward in time to get the desired gradient. A simpler and faster (but numerically less stable) method is to parametrize the control field in time and to calculate the gradient by a difference method,
\begin{eqnarray}
J(t) &\rightarrow& J(t_n)\equiv J_n, \; t_n=nh,\; n \in \left[ 0,N\right] \nonumber \\
N&=&\frac{t_f}{h}, \quad h...\mbox{grid spacing} \nonumber \\
\frac{\delta \mathcal{F}}{\delta J_n}&=&\frac{\mathcal{F}\left(J_0,...,J_n + \Delta J_n,... \right) -\mathcal{F}\left(  J_0,...,J_n,...\right) }{\Delta J_n} \nonumber \\
\frac{\delta \mathcal{F}}{\delta J} &\rightarrow& \left( \frac{\delta \mathcal{F}}{\delta J_0},\frac{\delta \mathcal{F}}{\delta J_1},...,\frac{\delta \mathcal{F}}{\delta J_N}\right),
\end{eqnarray}
which we  use here.
\par Utilizing the CG algorithm it is possible to optimize the parameters $J_n$ [see Fig.~\ref{fig:trap+opt}(e)]. A rapid initial and small final increase of the concurrence as well as a faster convergence of the density matrix elements were achieved by increasing the amplitude of the interdot coupling $J(t)$ at intermediate times [see Figs.~\ref{fig:trap+opt}(a)-\ref{fig:trap+opt}(d)]. It is interesting that the initial increase of concurrence was gathered at the  expense of purity at intermediate times [Fig.~\ref{fig:trap+opt}(f)].
However, final purity is significantly improved over the result from the initial guess for $J(t)$.
\begin{table}[h]
\begin{center}
\begin{tabular}{p{0.8cm}p{1cm}cccp{0.8cm}p{0.8cm}}
\hline
\hline
$\eta$				& 	$\hbar \omega_{sc}$		&	T					&	$t_{f}=t_{f,sc} / \omega_{sc}$				&	$\hbar \omega_c$	&	$B^{(1)}_{z}$	&	$B^{(2)}_{z}$ \\
\hline
0.08				&	1meV &	$50 \einheit{mK}$	&	$2.56 \einheit{ps}$&	$20 \einheit{meV}$	&	$1 \einheit{T}$		&	$1 \einheit{T}$ \\
\hline
\hline
\end{tabular}
\caption{Physical quantities used to calculate the data shown in Figs.~\ref{fig:trap+opt}(a)--\ref{fig:trap+opt}(f).}
\end{center}
\end{table}
\section{Conclusion and Outlook}\label{sec:conclusion-outlook}
It has been shown that entanglement of two electron spins in a semiconductor double quantum dot can be established and preserved by means of a tunable interdot voltage barrier.  The strategy of transforming the target (trapping) state into the ground state has been tested within a Markovian quantum master equation approach for a spin--boson coupling between the double qubit and its environment.  This analysis has shown that dephasing due to the environment modeled by two uncorrelated baths of harmonic oscillators can be suppressed.
When the strategy of making the target state to the ground state of the system is applied to an arbitrary maximally entangled state (Bell state), it has been shown that a detailed controllability  of the (symmetry) of the interdot coupling is required. 
However, an effective isotropic Heisenberg interaction between the two spins allows preparation of the selected Bell state $|\psi^\pm\rangle = \frac{1}{\sqrt{2}} (|1\rangle \otimes |0\rangle \pm |0\rangle \otimes |1\rangle)$.  Numerical results for the case of steering and trapping into this entangled state were given. 
An analysis of the dependence of the optimization technique on the temperature of the baths has shown that stable entanglement with respect to pure dephasing using experimentally realistic qubit--qubit couplings is possible over a broad temperature regime ($\lesssim 15 \einheit{K}$).
Furthermore, the control field was subjected to a CG optimization routine which leads to improved results: more rapid driving into the target Bell state and improved final purity and concurrence.
\section*{Acknowledgment}
The authors wish to acknowledge financial support of this work by FWF, Project Nos. P16317-N08 and P18829.
\bibliographystyle{apsrev}

\begin{thebibliography}{40}
\expandafter\ifx\csname natexlab\endcsname\relax\def\natexlab#1{#1}\fi
\expandafter\ifx\csname bibnamefont\endcsname\relax
  \def\bibnamefont#1{#1}\fi
\expandafter\ifx\csname bibfnamefont\endcsname\relax
  \def\bibfnamefont#1{#1}\fi
\expandafter\ifx\csname citenamefont\endcsname\relax
  \def\citenamefont#1{#1}\fi
\expandafter\ifx\csname url\endcsname\relax
  \def\url#1{\texttt{#1}}\fi
\expandafter\ifx\csname urlprefix\endcsname\relax\def\urlprefix{URL }\fi
\providecommand{\bibinfo}[2]{#2}
\providecommand{\eprint}[2][]{\url{#2}}

\bibitem[{\citenamefont{Nielsen and Chuang}(2002)}]{Niel02}
\bibinfo{author}{\bibfnamefont{M.}~\bibnamefont{Nielsen}} \bibnamefont{and}
  \bibinfo{author}{\bibfnamefont{I.}~\bibnamefont{Chuang}},
  \emph{\bibinfo{title}{Quantum computation and Quantum Information}}
  (\bibinfo{publisher}{Cambridge University Press}, \bibinfo{year}{2002}).

\bibitem[{\citenamefont{Shor}(1997)}]{Shor97}
\bibinfo{author}{\bibfnamefont{P.}~\bibnamefont{Shor}}, \bibinfo{journal}{SIAM
  J. Comput.} \textbf{\bibinfo{volume}{26}}, \bibinfo{pages}{1484}
  (\bibinfo{year}{1997}).

\bibitem[{\citenamefont{Loss and DiVincenzo}(1998)}]{DiVi98}
\bibinfo{author}{\bibfnamefont{D.}~\bibnamefont{Loss}} \bibnamefont{and}
  \bibinfo{author}{\bibfnamefont{D.}~\bibnamefont{DiVincenzo}},
  \bibinfo{journal}{Phys. Rev. A} \textbf{\bibinfo{volume}{57}},
  \bibinfo{pages}{120} (\bibinfo{year}{1998}).

\bibitem[{\citenamefont{Yoran and Reznik}(2003)}]{Yor03}
\bibinfo{author}{\bibfnamefont{N.}~\bibnamefont{Yoran}} \bibnamefont{and}
  \bibinfo{author}{\bibfnamefont{B.}~\bibnamefont{Reznik}},
  \bibinfo{journal}{Phys. Rev. Letters} \textbf{\bibinfo{volume}{91}},
  \bibinfo{pages}{037903} (\bibinfo{year}{2003}).

\bibitem[{\citenamefont{Gershenfeld and Chuang}(1997)}]{Ger97}
\bibinfo{author}{\bibfnamefont{N.}~\bibnamefont{Gershenfeld}} \bibnamefont{and}
  \bibinfo{author}{\bibfnamefont{I.}~\bibnamefont{Chuang}},
  \bibinfo{journal}{Science} \textbf{\bibinfo{volume}{275}},
  \bibinfo{pages}{350} (\bibinfo{year}{1997}).

\bibitem[{\citenamefont{Orlando et~al.}(1999)\citenamefont{Orlando, Mooij,
  Tian, derWal, Levitov, Lloyd, and Mazo}}]{Orl99}
\bibinfo{author}{\bibfnamefont{T.}~\bibnamefont{Orlando}},
  \bibinfo{author}{\bibfnamefont{J.~E.} \bibnamefont{Mooij}},
  \bibinfo{author}{\bibfnamefont{L.}~\bibnamefont{Tian}},
  \bibinfo{author}{\bibfnamefont{C.~H.~V.} \bibnamefont{derWal}},
  \bibinfo{author}{\bibfnamefont{L.}~\bibnamefont{Levitov}},
  \bibinfo{author}{\bibfnamefont{S.}~\bibnamefont{Lloyd}}, \bibnamefont{and}
  \bibinfo{author}{\bibfnamefont{J.~J.} \bibnamefont{Mazo}},
  \bibinfo{journal}{Phys. Rev. B} \textbf{\bibinfo{volume}{60}},
  \bibinfo{pages}{15398} (\bibinfo{year}{1999}).

\bibitem[{\citenamefont{Makhlin et~al.}(1999)\citenamefont{Makhlin, Schoen, and
  Shnirman}}]{Mak99}
\bibinfo{author}{\bibfnamefont{Y.}~\bibnamefont{Makhlin}},
  \bibinfo{author}{\bibfnamefont{G.}~\bibnamefont{Schoen}}, \bibnamefont{and}
  \bibinfo{author}{\bibfnamefont{A.}~\bibnamefont{Shnirman}},
  \bibinfo{journal}{Nature} \textbf{\bibinfo{volume}{398}},
  \bibinfo{pages}{305} (\bibinfo{year}{1999}).

\bibitem[{\citenamefont{Cirac and Zoller}(1995)}]{Cir95}
\bibinfo{author}{\bibfnamefont{J.}~\bibnamefont{Cirac}} \bibnamefont{and}
  \bibinfo{author}{\bibfnamefont{P.}~\bibnamefont{Zoller}},
  \bibinfo{journal}{Phys. Rev. Letters} \textbf{\bibinfo{volume}{74}},
  \bibinfo{pages}{4091} (\bibinfo{year}{1995}).

\bibitem[{\citenamefont{Preskill}(1998)}]{Pres98}
\bibinfo{author}{\bibfnamefont{J.}~\bibnamefont{Preskill}},
  \bibinfo{journal}{Proc. R. Soc. Lond. A} \textbf{\bibinfo{volume}{454}},
  \bibinfo{pages}{385} (\bibinfo{year}{1998}).

\bibitem[{\citenamefont{Lidar et~al.}(1998)\citenamefont{Lidar, Chuang, and
  Whaley}}]{Lid98}
\bibinfo{author}{\bibfnamefont{D.}~\bibnamefont{Lidar}},
  \bibinfo{author}{\bibfnamefont{I.}~\bibnamefont{Chuang}}, \bibnamefont{and}
  \bibinfo{author}{\bibfnamefont{K.}~\bibnamefont{Whaley}},
  \bibinfo{journal}{Phys. Rev. Letters} \textbf{\bibinfo{volume}{81}},
  \bibinfo{pages}{2594} (\bibinfo{year}{1998}).

\bibitem[{\citenamefont{Lidar et~al.}(1999)\citenamefont{Lidar, Bacon, and
  Whaley}}]{Lid99}
\bibinfo{author}{\bibfnamefont{D.}~\bibnamefont{Lidar}},
  \bibinfo{author}{\bibfnamefont{D.}~\bibnamefont{Bacon}}, \bibnamefont{and}
  \bibinfo{author}{\bibfnamefont{K.}~\bibnamefont{Whaley}},
  \bibinfo{journal}{Phys. Rev. Letters} \textbf{\bibinfo{volume}{82}},
  \bibinfo{pages}{4556} (\bibinfo{year}{1999}).

\bibitem[{\citenamefont{Viola and Lloyd}(1998)}]{Viol98}
\bibinfo{author}{\bibfnamefont{L.}~\bibnamefont{Viola}} \bibnamefont{and}
  \bibinfo{author}{\bibfnamefont{S.}~\bibnamefont{Lloyd}},
  \bibinfo{journal}{Phys. Rev. A} \textbf{\bibinfo{volume}{58}},
  \bibinfo{pages}{2733} (\bibinfo{year}{1998}).

\bibitem[{\citenamefont{Viola et~al.}(1999)\citenamefont{Viola, Knill, and
  Lloyd}}]{Viol99}
\bibinfo{author}{\bibfnamefont{L.}~\bibnamefont{Viola}},
  \bibinfo{author}{\bibfnamefont{E.}~\bibnamefont{Knill}}, \bibnamefont{and}
  \bibinfo{author}{\bibfnamefont{S.}~\bibnamefont{Lloyd}},
  \bibinfo{journal}{Phys. Rev. Letters} \textbf{\bibinfo{volume}{82}},
  \bibinfo{pages}{2417} (\bibinfo{year}{1999}).

\bibitem[{\citenamefont{Poetz et~al.}(2006)\citenamefont{Poetz, Goritschnig,
  and H.Jirari}}]{Poetz1}
\bibinfo{author}{\bibfnamefont{W.}~\bibnamefont{Poetz}},
  \bibinfo{author}{\bibfnamefont{A.}~\bibnamefont{Goritschnig}},
  \bibnamefont{and} \bibinfo{author}{\bibnamefont{H.Jirari}},
  \bibinfo{journal}{Proc. 5$^{th}$ MATHMOD, Feb. 8-10, ARGESIM Report no.~30,
  Vol. 1 p.~87, and Physical Modeling pp~9-1 to~9-10}  (\bibinfo{year}{2006}).

\bibitem[{\citenamefont{Elzerman et~al.}(2005)}]{Elz}
\bibinfo{author}{\bibfnamefont{J.}~\bibnamefont{Elzerman}}
  \bibnamefont{et~al.}, \emph{\bibinfo{title}{Semiconductor few-electron
  quantum dots as spin qubits}} (\bibinfo{publisher}{Springer},
  \bibinfo{year}{2005}).

\bibitem[{\citenamefont{Burkard et~al.}(1999)\citenamefont{Burkard, Loss, and
  DiVincenzo}}]{Burk99}
\bibinfo{author}{\bibfnamefont{G.}~\bibnamefont{Burkard}},
  \bibinfo{author}{\bibfnamefont{D.}~\bibnamefont{Loss}}, \bibnamefont{and}
  \bibinfo{author}{\bibfnamefont{D.~P.} \bibnamefont{DiVincenzo}},
  \bibinfo{journal}{Phys. Rev. B} \textbf{\bibinfo{volume}{59}},
  \bibinfo{pages}{2070} (\bibinfo{year}{1999}).

\bibitem[{\citenamefont{Trif et~al.}(2007)\citenamefont{Trif, Golovach, and
  Loss}}]{Trif07}
\bibinfo{author}{\bibfnamefont{M.}~\bibnamefont{Trif}},
  \bibinfo{author}{\bibfnamefont{V.~N.} \bibnamefont{Golovach}},
  \bibnamefont{and} \bibinfo{author}{\bibfnamefont{D.}~\bibnamefont{Loss}},
  \bibinfo{journal}{Phys. Rev. B} \textbf{\bibinfo{volume}{75}},
  \bibinfo{pages}{085307} (\bibinfo{year}{2007}).

\bibitem[{\citenamefont{Carmichael}(2002)}]{Car02}
\bibinfo{author}{\bibfnamefont{H.}~\bibnamefont{Carmichael}},
  \emph{\bibinfo{title}{Statistical Methods in Quantum Optics 1: Master
  Equations and Fokker--Planck Equations}} (\bibinfo{publisher}{Springer},
  \bibinfo{year}{2002}).

\bibitem[{\citenamefont{Breuer and Petruccione}(2003)}]{Breu03}
\bibinfo{author}{\bibfnamefont{H.}~\bibnamefont{Breuer}} \bibnamefont{and}
  \bibinfo{author}{\bibfnamefont{F.}~\bibnamefont{Petruccione}},
  \emph{\bibinfo{title}{The theory of open quantum systems}}
  (\bibinfo{publisher}{Oxford}, \bibinfo{year}{2003}).

\bibitem[{\citenamefont{Scully and Zubairy}(1997)}]{Scully}
\bibinfo{author}{\bibfnamefont{M.~O.} \bibnamefont{Scully}} \bibnamefont{and}
  \bibinfo{author}{\bibfnamefont{M.~S.} \bibnamefont{Zubairy}},
  \emph{\bibinfo{title}{Quantum Optics}} (\bibinfo{publisher}{Cambridge
  University Press}, \bibinfo{year}{1997}).

\bibitem[{\citenamefont{Taylor et~al.}(2006)\citenamefont{Taylor, Petta,
  Johnson, Yacoby, Marcus, and Lukin}}]{Tay06}
\bibinfo{author}{\bibfnamefont{J.}~\bibnamefont{Taylor}},
  \bibinfo{author}{\bibfnamefont{J.~R.} \bibnamefont{Petta}},
  \bibinfo{author}{\bibfnamefont{A.~C.} \bibnamefont{Johnson}},
  \bibinfo{author}{\bibfnamefont{A.}~\bibnamefont{Yacoby}},
  \bibinfo{author}{\bibfnamefont{C.~M.} \bibnamefont{Marcus}},
  \bibnamefont{and} \bibinfo{author}{\bibfnamefont{M.~D.} \bibnamefont{Lukin}},
  \bibinfo{journal}{arXiv:cond-mat/0602470 v1}  (\bibinfo{year}{2006}).

\bibitem[{\citenamefont{Golovach et~al.}(2004)\citenamefont{Golovach,
  Khaetskii, and Loss}}]{Gol04}
\bibinfo{author}{\bibfnamefont{V.}~\bibnamefont{Golovach}},
  \bibinfo{author}{\bibfnamefont{A.}~\bibnamefont{Khaetskii}},
  \bibnamefont{and} \bibinfo{author}{\bibfnamefont{D.}~\bibnamefont{Loss}},
  \bibinfo{journal}{Phys. Rev. Letters} \textbf{\bibinfo{volume}{93}},
  \bibinfo{pages}{016601} (\bibinfo{year}{2004}).

\bibitem[{\citenamefont{Mozyrsky et~al.}(2002)\citenamefont{Mozyrsky, Kogan,
  Gorshkov, and Berman}}]{Moz02}
\bibinfo{author}{\bibfnamefont{D.}~\bibnamefont{Mozyrsky}},
  \bibinfo{author}{\bibfnamefont{S.}~\bibnamefont{Kogan}},
  \bibinfo{author}{\bibfnamefont{V.~N.} \bibnamefont{Gorshkov}},
  \bibnamefont{and} \bibinfo{author}{\bibfnamefont{G.~P.}
  \bibnamefont{Berman}}, \bibinfo{journal}{Phys. Rev. B}
  \textbf{\bibinfo{volume}{65}}, \bibinfo{pages}{245213}
  (\bibinfo{year}{2002}).

\bibitem[{\citenamefont{Yu and Eberly}(2002)}]{Yu02}
\bibinfo{author}{\bibfnamefont{T.}~\bibnamefont{Yu}} \bibnamefont{and}
  \bibinfo{author}{\bibfnamefont{J.}~\bibnamefont{Eberly}},
  \bibinfo{journal}{Phys. Rev. B} \textbf{\bibinfo{volume}{66}},
  \bibinfo{pages}{193306} (\bibinfo{year}{2002}).

\bibitem[{\citenamefont{Hu and Sarma}(2006)}]{Hu06}
\bibinfo{author}{\bibfnamefont{X.}~\bibnamefont{Hu}} \bibnamefont{and}
  \bibinfo{author}{\bibfnamefont{S.~D.} \bibnamefont{Sarma}},
  \bibinfo{journal}{Phys. Rev. Letters} \textbf{\bibinfo{volume}{96}},
  \bibinfo{pages}{100501} (\bibinfo{year}{2006}).

\bibitem[{\citenamefont{Mitin et~al.}(1999)\citenamefont{Mitin, Kochelap, and
  Stroscio}}]{Mit99}
\bibinfo{author}{\bibfnamefont{V.}~\bibnamefont{Mitin}},
  \bibinfo{author}{\bibfnamefont{V.}~\bibnamefont{Kochelap}}, \bibnamefont{and}
  \bibinfo{author}{\bibfnamefont{M.}~\bibnamefont{Stroscio}},
  \emph{\bibinfo{title}{Quantum Heterostructures}}
  (\bibinfo{publisher}{Cambridge University Press}, \bibinfo{year}{1999}).

\bibitem[{\citenamefont{Leggett}(1987)}]{Leg87}
\bibinfo{author}{\bibfnamefont{A.}~\bibnamefont{Leggett}},
  \bibinfo{journal}{Reviews of Modern Physics} \textbf{\bibinfo{volume}{59}}
  (\bibinfo{year}{1987}).

\bibitem[{\citenamefont{Weiss}(1999)}]{Weiss99}
\bibinfo{author}{\bibfnamefont{U.}~\bibnamefont{Weiss}},
  \emph{\bibinfo{title}{Quantum dissipative systems}}
  (\bibinfo{publisher}{World Scientific}, \bibinfo{year}{1999}).

\bibitem[{wol()}]{wolfram}
\urlprefix\url{http://functions.wolfram.com/}.

\bibitem[{\citenamefont{Brandes and Vorrath}(2002)}]{Brand02}
\bibinfo{author}{\bibfnamefont{T.}~\bibnamefont{Brandes}} \bibnamefont{and}
  \bibinfo{author}{\bibfnamefont{T.}~\bibnamefont{Vorrath}},
  \bibinfo{journal}{Phys. Rev. B} \textbf{\bibinfo{volume}{66}},
  \bibinfo{pages}{075341} (\bibinfo{year}{2002}).

\bibitem[{\citenamefont{Westfahl et~al.}(2004)\citenamefont{Westfahl, Caldeira,
  Medeiros-Ribeiro, and Cerro}}]{West04}
\bibinfo{author}{\bibfnamefont{H.}~\bibnamefont{Westfahl}},
  \bibinfo{author}{\bibfnamefont{A.~O.} \bibnamefont{Caldeira}},
  \bibinfo{author}{\bibfnamefont{G.}~\bibnamefont{Medeiros-Ribeiro}},
  \bibnamefont{and} \bibinfo{author}{\bibfnamefont{M.}~\bibnamefont{Cerro}},
  \bibinfo{journal}{Phys. Rev. B} \textbf{\bibinfo{volume}{70}},
  \bibinfo{pages}{195320} (\bibinfo{year}{2004}).

\bibitem[{\citenamefont{Palma et~al.}(1996)\citenamefont{Palma, Suominen, and
  Ekert}}]{Pal96}
\bibinfo{author}{\bibfnamefont{G.}~\bibnamefont{Palma}},
  \bibinfo{author}{\bibfnamefont{K.}~\bibnamefont{Suominen}}, \bibnamefont{and}
  \bibinfo{author}{\bibfnamefont{A.}~\bibnamefont{Ekert}},
  \bibinfo{journal}{Proc. R. Soc. Lond. A} \textbf{\bibinfo{volume}{452}},
  \bibinfo{pages}{567} (\bibinfo{year}{1996}).

\bibitem[{\citenamefont{Buettiker}(2001)}]{Buett01}
\bibinfo{author}{\bibfnamefont{M.}~\bibnamefont{Buettiker}},
  \bibinfo{journal}{arXiv:cond-mat/0106149 v1}  (\bibinfo{year}{2001}).

\bibitem[{\citenamefont{Reina et~al.}(2002)\citenamefont{Reina, Quiroga, and
  Johnson}}]{Rei02}
\bibinfo{author}{\bibfnamefont{J.}~\bibnamefont{Reina}},
  \bibinfo{author}{\bibfnamefont{L.}~\bibnamefont{Quiroga}}, \bibnamefont{and}
  \bibinfo{author}{\bibfnamefont{N.}~\bibnamefont{Johnson}},
  \bibinfo{journal}{Phys. Rev. A} \textbf{\bibinfo{volume}{65}},
  \bibinfo{pages}{032326} (\bibinfo{year}{2002}).

\bibitem[{\citenamefont{Wootters}(1998)}]{Wo98}
\bibinfo{author}{\bibfnamefont{W.}~\bibnamefont{Wootters}},
  \bibinfo{journal}{Phys. Rev. Letters} \textbf{\bibinfo{volume}{80}},
  \bibinfo{pages}{2245} (\bibinfo{year}{1998}).

\bibitem[{\citenamefont{Schwabl}(2004)}]{Schwabl}
\bibinfo{author}{\bibfnamefont{F.}~\bibnamefont{Schwabl}},
  \emph{\bibinfo{title}{Statistische Mechanik}} (\bibinfo{publisher}{Springer},
  \bibinfo{year}{2004}).

\bibitem[{\citenamefont{Vetterling et~al.}(1995)\citenamefont{Vetterling,
  Teukolsky, Press, and Flannery}}]{NrC}
\bibinfo{author}{\bibfnamefont{W.}~\bibnamefont{Vetterling}},
  \bibinfo{author}{\bibfnamefont{S.}~\bibnamefont{Teukolsky}},
  \bibinfo{author}{\bibfnamefont{W.}~\bibnamefont{Press}}, \bibnamefont{and}
  \bibinfo{author}{\bibfnamefont{B.}~\bibnamefont{Flannery}},
  \emph{\bibinfo{title}{Numerical Recipes in C}} (\bibinfo{publisher}{Cambridge
  University Press}, \bibinfo{year}{1995}).

\bibitem[{\citenamefont{Jirari and Poetz}(2005)}]{Poetz3}
\bibinfo{author}{\bibfnamefont{H.}~\bibnamefont{Jirari}} \bibnamefont{and}
  \bibinfo{author}{\bibfnamefont{W.}~\bibnamefont{Poetz}},
  \bibinfo{journal}{Phys. Rev. A} \textbf{\bibinfo{volume}{72}},
  \bibinfo{pages}{013409} (\bibinfo{year}{2005}).

\bibitem[{\citenamefont{Jirari and Poetz}(2006)}]{Poetz2}
\bibinfo{author}{\bibfnamefont{H.}~\bibnamefont{Jirari}} \bibnamefont{and}
  \bibinfo{author}{\bibfnamefont{W.}~\bibnamefont{Poetz}},
  \bibinfo{journal}{Phys. Rev. A} \textbf{\bibinfo{volume}{74}},
  \bibinfo{pages}{022306} (\bibinfo{year}{2006}).

\bibitem[{\citenamefont{Benatti et~al.}(2003)\citenamefont{Benatti,
  R.Floreanini, and Piani}}]{Ben03}
\bibinfo{author}{\bibfnamefont{F.}~\bibnamefont{Benatti}},
  \bibinfo{author}{\bibnamefont{R.Floreanini}}, \bibnamefont{and}
  \bibinfo{author}{\bibfnamefont{M.}~\bibnamefont{Piani}},
  \bibinfo{journal}{Phys. Rev. A} \textbf{\bibinfo{volume}{67}},
  \bibinfo{pages}{042110} (\bibinfo{year}{2003}).

\end{thebibliography}
\begin{appendix}
\section{Calculation of the correlation functions}\label{app:corrf}
The bath operators of Eq.~\eqref{corrf0} in the interaction picture read
\begin{equation}
\tilde \Gamma_i(t)=\sum\limits_k{g_k^{(i)} \left( e^{-i \omega_k^{(i)} t}\tilde b_k^{(i)}(0)+e^{i \omega_k^{(i)} t} \tilde b_k^{(i)\dag}(0) \right)}.
\end{equation}
Because $\rho_R$ [see Eq.~\eqref{rho_R}] is diagonal in the bosonic occupation numbers, we get
\begin{eqnarray}
\lefteqn{\left\langle {\tilde \Gamma^{(i)}(t) \tilde \Gamma^{(i)}(t') } \right\rangle _R=\operatorname{tr_R}\left\lbrace \tilde \Gamma^{(i)}(t) \tilde \Gamma^{(i)}(t') \tilde \rho_R^{(i)} \right\rbrace=} \label{exp3} \\
&=& \sum_k{ \left( g_{k}^{(i)}\right)^2 \left[ e^{-i\omega_k^{(i)}(t-t')} \left( \bar n_k^{(i)}+1\right)  + e^{i\omega_k^{(i)}(t-t')} \bar n_k^{(i)} \right]}, \nonumber 
\end{eqnarray}
where $\bar n_k^{(i)}$ denotes the expectation value of the particle number operator $\tilde b_k^{(i)\dag} \tilde b_k^{(i)}$.
Now, define the \textit{spectral density}~\cite{Leg87, Weiss99}
\begin{equation}
J(\omega)=\frac{\pi}{2} \sum_k{\left( g_k^{(i)}\right) ^2 \delta \left( \omega - \omega_{k}^{(i)}\right) }. \label{spectral1}
\end{equation}
By using Eq.~\eqref{spectral1}, we can reformulate Eq.~\eqref{exp3} as
\begin{eqnarray}
\left\langle {\tilde \Gamma _i(t) \tilde \Gamma _i(t') } \right\rangle _B &=& \frac{2}{\pi}\int\limits_0^\infty  {d\omega J(\omega )\left[ {\left( {\bar n^{(i)}(\omega) + 1} \right)e^{-i\omega(t-t')}}\right. +}  \nonumber \\
&&+\left. \bar n^{(i)}(\omega) e^{i\omega (t-t')} \right] \label{spectral2}
\end{eqnarray}
If we use the continuum limit [Eq.~\eqref{ohmic}] and
\begin{eqnarray}
\bar n^{(i)}(\omega)&=&\frac{1}{e^{\beta \hbar \omega}-1}, \\
\beta &=&\frac{1}{k_B T},
\end{eqnarray}
we can write Eq.~\eqref{spectral2} as
\begin{eqnarray}
\left\langle {\tilde \Gamma _i(t) \tilde \Gamma _i(t') } \right\rangle _B &=& \frac{2\eta}{\pi}\int\limits_0^\infty  {d\omega \left\lbrace \omega e^{-\frac{\omega}{\omega_c}} \left[ {\frac{e^{\beta \hbar \omega}}{e^{\beta \hbar \omega}-1} e^{-i\omega(t-t')}} + \right. \right.} \nonumber \\
&& + \left. \left. \frac{1}{e^{\beta \hbar \omega}-1} e^{i\omega (t-t')} \right]\right\rbrace.
\end{eqnarray}
This integration can be done analytically yielding Eq.~\eqref{corrf1}.
\section{Dependence of the dynamics on the coupling strength $\eta$}\label{app:damping}
\begin{figure}[h]
\begin{tabular*}{8.5cm}{ll}
(a) & (b) \\
\includegraphics[width=4.25cm]{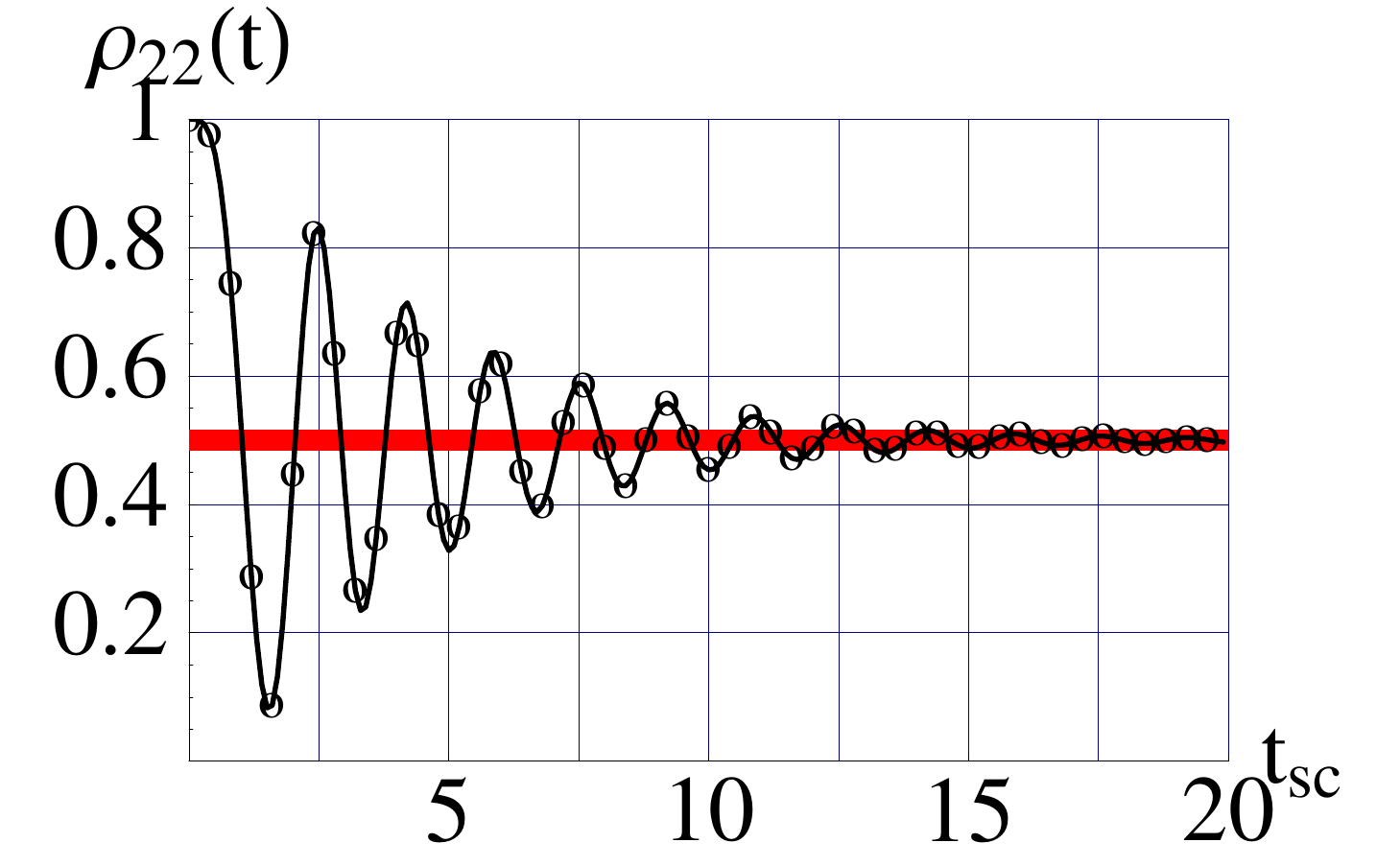} & \includegraphics[width=4.25cm]{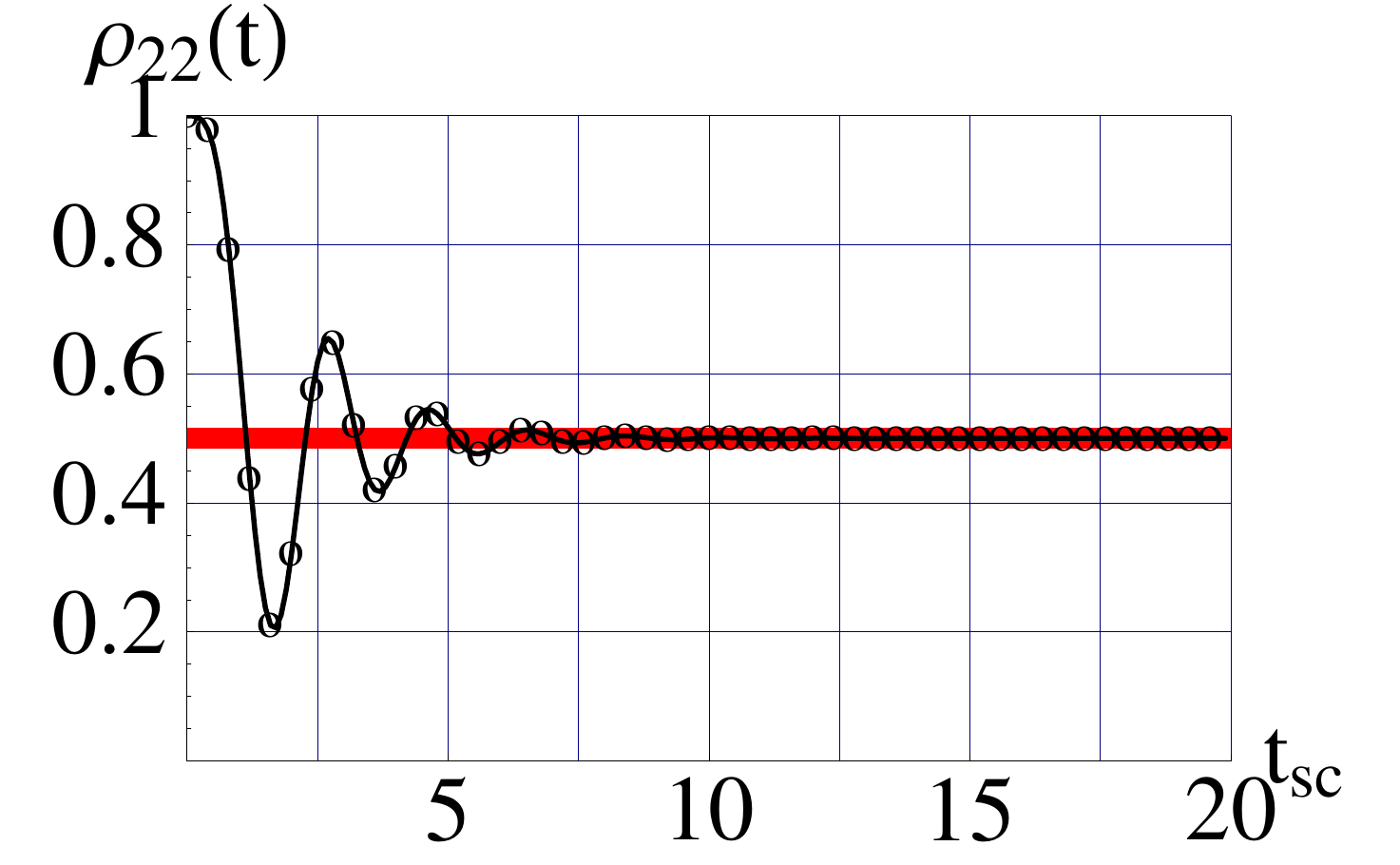} \\
(c) & (d) \\
\includegraphics[width=4.25cm]{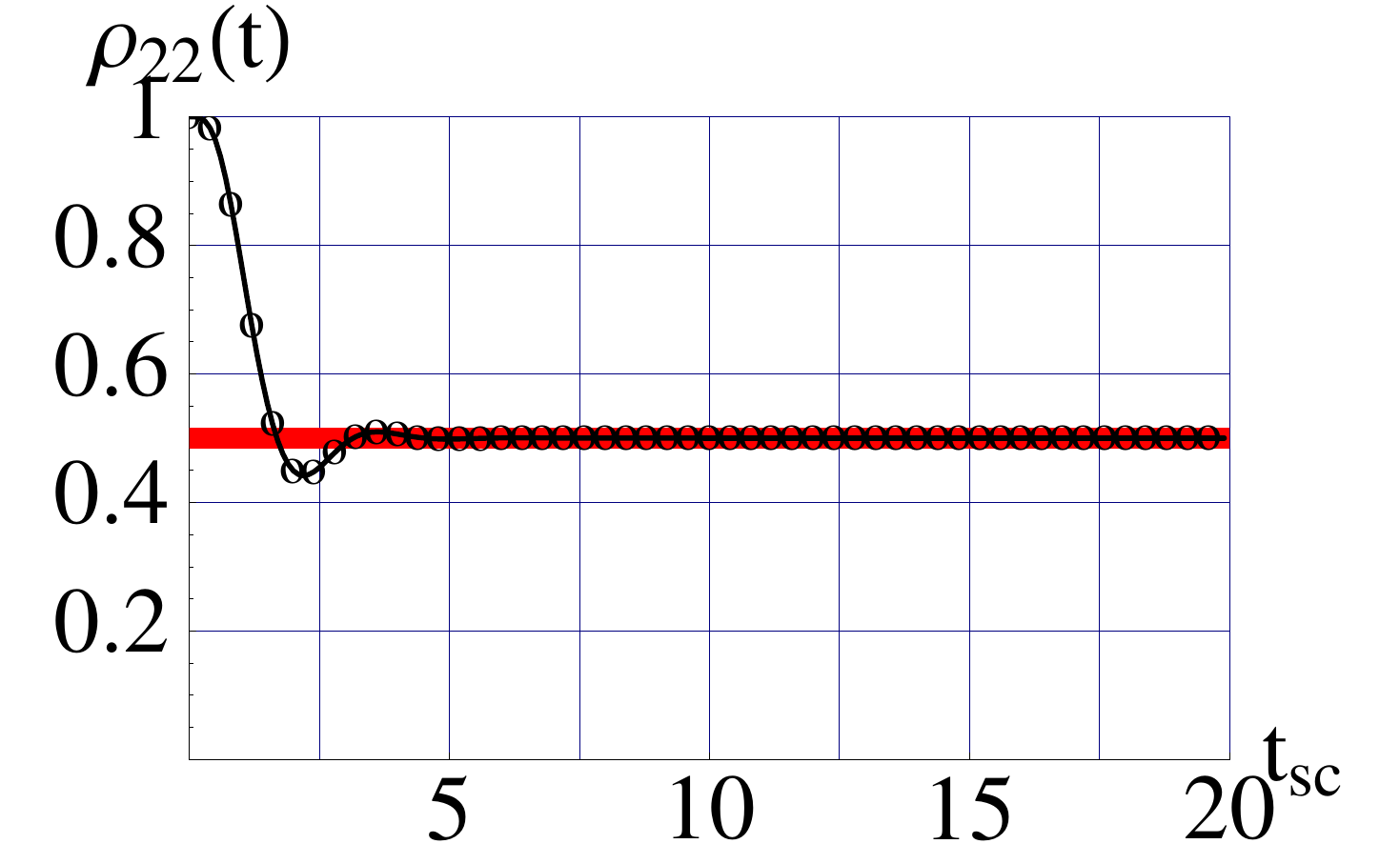} & \includegraphics[width=4.25cm]{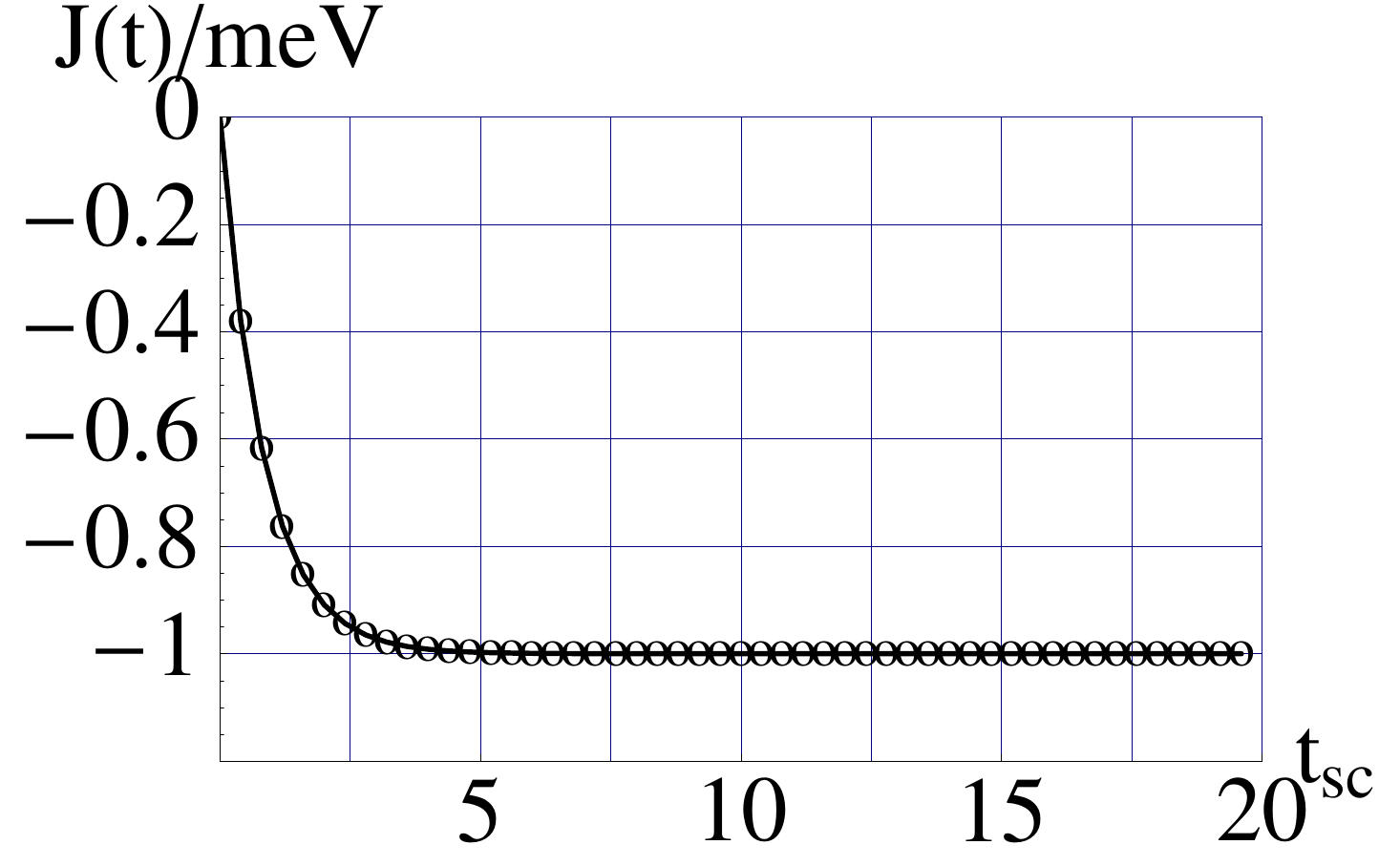}
\end{tabular*}
\caption{\label{fig:dyn_eta}Red (thick solid) lines indicate the desired values of the density matrix element. The diagonal density matrix element $\rho_{22}$ vs time are shown for (a) $\eta=0.02$, (b) $0.05$, and (c) $0.10$. The control field $J(t)$ is given in (d). For the remaining physical quantities, we used the same values as in Table~\ref{tab2}.}
\end{figure}

In Figs.~\ref{fig:dyn_eta}(a)--\ref{fig:dyn_eta}(c) the oscillations of the diagonal density matrix element $\rho_{22}$ with respect to a variation of the coupling strength $\eta$ are shown. The control field $J(t)$ is shown in Fig.~\ref{fig:dyn_eta}(d).
As can be seen, there is an analogy to classical mechanics. For $\eta=0.02\;\mbox{and}\;\eta=0.05$, the system is ``underdamped'', whereas for $\eta=0.10$ the system is quite close to the ``critically damped'' case. The equilibrium position corresponds to the desired state $|\psi^+\rangle \langle\psi^+|$ [see Eq.~\eqref{di_state}]. Unfortunately, we cannot approach the ``overdamped'' parameter regime because of positivity problems. These problems, e.g., $P(t) > 1$, result from straining the Born--Markov approximation, \textit{i.e.}, the system--environment interaction becomes too strong.~\cite{Ben03}
\end{appendix}

\end{document}